\documentclass[10pt,journal,compsoc]{IEEEtran}
\ifCLASSOPTIONcompsoc
  \usepackage[nocompress]{cite}
\else
  \usepackage{cite}
\fi
\ifCLASSINFOpdf
\else
\fi
\usepackage{amsmath}

\usepackage{xspace}
\usepackage{amsfonts}
\usepackage{bm}
\usepackage{multirow}
\usepackage{orcidlink}

\newcommand{\sys}{NNInv\xspace}
\definecolor{mred}{rgb}{.80,.12,.30}

\begin{document}
%
\title{UnProjection: Leveraging Inverse-Projections for Visual Analytics of High-Dimensional Data}
%
%
%
%
\author{Mateus Espadoto\,\orcidlink{0000-0002-1922-4309}, Gabriel Appleby\,\orcidlink{0000-0003-2436-2121}, Ashley Suh\,\orcidlink{0000-0001-6513-8447}, Dylan Cashman\,\orcidlink{0000-0003-4853-5701}, Mingwei Li\,\orcidlink{0000-0002-0457-8091},\\ Carlos Scheidegger, Erik W Anderson\,\orcidlink{0000-0002-0334-8497}, Remco Chang\,\orcidlink{0000-0002-6484-6430}, Alexandru C Telea\,\orcidlink{0000-0003-0750-0502}
\IEEEcompsocitemizethanks{
\IEEEcompsocthanksitem M. Espadoto is with the University of University of Sao Paulo.
\IEEEcompsocthanksitem G. Appleby, A. Suh, and R. Chang are with Tufts University.
\IEEEcompsocthanksitem D. Cashman was with Tufts University. He is now with Novartis.
\IEEEcompsocthanksitem M. Li, and C. Scheidegger are with the University of Arizona.
\IEEEcompsocthanksitem E. Anderson was with Northeastern University. He is now with Novartis.
\IEEEcompsocthanksitem A. Telea is with the University of Utrecht.}
\thanks{Manuscript received April 19, 2005; revised August 26, 2015.}}

\IEEEtitleabstractindextext{%
\begin{abstract}
Projection techniques are often used to visualize high-dimensional data, allowing users to better understand the overall structure of multi-dimensional spaces on a 2D screen. Although many such methods exist, comparably little work has been done on generalizable methods of inverse-projection -- the process of mapping the projected points, or more generally, the projection space back to the original high-dimensional space. 
In this paper we present \sys, a deep learning technique with the ability to approximate the inverse of any projection or mapping. 
\sys learns to reconstruct high-dimensional data from any arbitrary point on a 2D projection space, giving users the ability to interact with the learned high-dimensional representation in a visual analytics system. 
We provide an analysis of the parameter space of \sys, and offer guidance in selecting these parameters.
We extend validation of the effectiveness of \sys through a series of quantitative and qualitative analyses. 
We then demonstrate the method's utility by applying it to three visualization tasks: interactive instance interpolation, classifier agreement, and gradient visualization.
\end{abstract}

\begin{IEEEkeywords}
Multidimensional data, Multidimensional projection, Inverse-projection, Back-projection.
\end{IEEEkeywords}}

\maketitle

\IEEEdisplaynontitleabstractindextext

%
\IEEEpeerreviewmaketitle
\IEEEraisesectionheading{\section{Introduction}}
%
\label{sec:introduction}
\IEEEPARstart{H}{igh-dimensional} data is created at unprecedented rates by scientific fields as diverse as information technology, bioinformatics, and astronomy\cite{buhlmann:2011:stats_for_high_dim}. As a result, there is a growing need for visualization and interaction methods for high-dimensional data.  A common choice is to project the high-dimensional data to two dimensions using methods such as t-SNE\cite{maaten:2008:tsne}, PCA\cite{pearson:1901:pca}, LLE\cite{roweis:2000:lle}, or UMAP\cite{mcinnes:2018:umap}, among the many other existing options\cite{espadoto19}.  
Projections allow better insight into the overall structure of data and can be enriched by interactions that allow users to reason about the corresponding high-dimensional data by selecting, brushing, and querying the 2D scatterplots they create. For example, t-SNE has allowed computational biologists to investigate human genetic data, revealing otherwise obfuscated population stratification\cite{Li:2017:application_of_tsne}. However, any projection technique will create errors when mapping complex and high-dimensional datasets to a low number of dimensions\cite{martins14,nonato18}. Moreover, projections are often complex algorithms, so the way they map the high-dimensional data to low-dimensional space can be difficult for users to fully interpret. As such, additional mechanisms need to be complement projections to empower users to better explore the high-dimensional data.
Recently attention has turned to \textit{inverse-projection}, a process that allows one to compute the inverse mapping from the projection space back to the original high-dimensional data space\cite{amorim:2012:ilamp}. Also called back-projections, these methods help users to explore projections by allowing a user to interactively query the projection space to find high-dimensional data points. These points correspond to specific locations in the low-dimensional projection. Inverse-projections are also instrumental in explaining the decision boundaries of machine learning classifiers\cite{rodrigues:2019:classifier_boundaries} and data augmentation scenarios\cite{amorim:2012:ilamp}. In contrast to the many existing projection techniques\cite{espadoto19}, only a handful of inverse projection algorithms exist, including iLAMP\cite{amorim:2012:ilamp} and its extension that uses radial basis functions (RBFs)\cite{amorim:2015:rbf}. Algorithms like iLAMP and RBF are quite slow, and have multiple free parameters, making it hard to use them in interactive data exploration scenarios\cite{rodrigues:2019:classifier_boundaries}.
In this paper, we present \sys, a technique for computing the inverse of any projection using a deep learning approach. Our idea is inspired by the recent work of Espadoto \emph{et al.}\cite{espadoto:2020:innp} that demonstrates that deep learning can learn to imitate the style of any projection technique, and is parametric and stable to data changes (and thereby offers out-of-sample capabilities). Following their approach, we show that \sys is a scalable, robust, high-quality inverse projection method which supports multiple applications. 
Using \sys, we introduce three use cases across a number of well-known datasets to illustrate how the use of inverse-projection can improve the user's interaction, exploration, and understanding of high-dimensional data in a 2D visualization. Additionally, we provide an evaluation of iLAMP, RBF, and \sys in terms of scalability and accuracy.
To this end, we provide a novel visualization for evaluating the joint quality of a pair of inverse-projection and direct-projection methods.
We make a point of studying the \sys inverse-projection method on two synthetic datasets with well-known topology (\emph{i.e.}, a 3D sphere dataset and a 3D swissroll dataset), allowing us to illustrate the behaviors of the learned inverse function. 
Applications for this work are numerous.  First, we use inverse-projections to explore the ``empty'' spaces in a 2D projection of high-dimensional data. While the user interactively brushes such spaces, high-dimensional instances corresponding to the visited 2D points are synthesized and displayed, thus allowing one to form a better mental map of how the 2D image represents the entire high-dimensional data \emph{space}, beyond how a 2D scatterplot represents a high-dimensional \emph{dataset}. 
Second, we present a visualization of the decision boundary of an ensemble classifier. Visualizing cluster boundaries can help users  see patterns within the data and the behaviors of the classifiers (see the classifier comparison by Scikit-Learn\,\cite{scikitboundary}). We show that the use of an inverse projection method such as \sys makes it possible to visualize this important information with high-dimensional data (see Fig.~\ref{fig:interpolation_table}). Finally, we introduce a gradient map visualization to help users find projection artifacts. This method highlights regions where the projection shrinks and expands the relationships between points by visualizing the rate of change between the learned 2D embedding and the original high-dimensional space. We show that  \sys provides an alternative approach to helping the user ``see'' the high-dimensional space in 2D.
In summary, the main contributions of this paper are:
\begin{itemize}
    \item A deep learning approach to inverse-projection, which is fast enough to be used at scale.
    \item A comparison to existing inverse-projection methods.
    \item An exploration of the behavior of \sys on datasets with well-known topology.
    \item Two novel visualizations for evaluating inverse-projection methods.
    \item A showcase of visual exploration techniques enabled by inverse-projection.
\end{itemize}

\begin{figure}[t]
    \centering
    \includegraphics[width=0.9\linewidth]{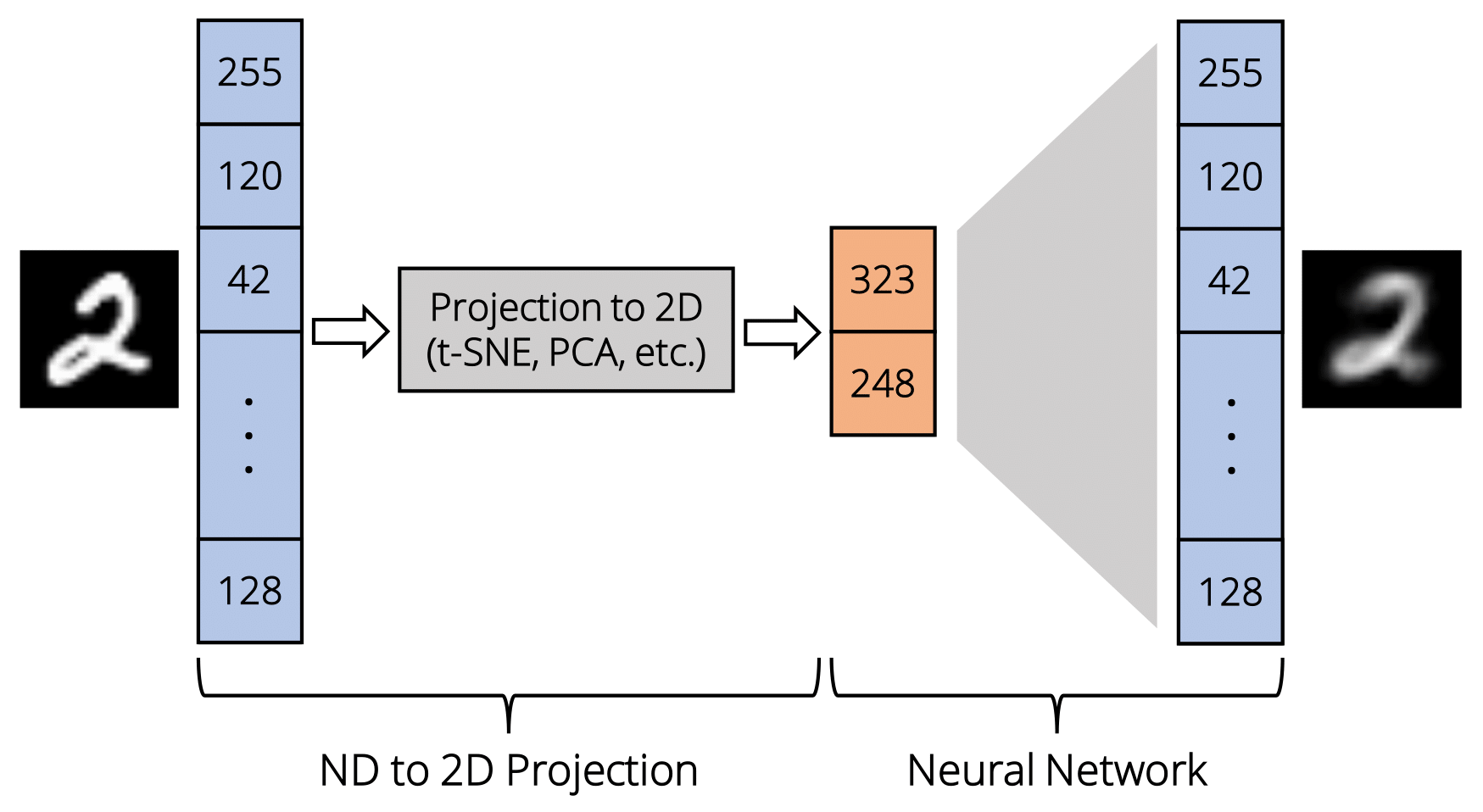} 
    \caption
    { End-to-end pipeline of direct and inverse projections. A high-dimensional vector representing an image of the number two is fed into a projection technique.
        The orange numbers are the 2D projection of this vector. The inverse projection \sys takes this 2D representation and yields a high-dimensional vector corresponding to it. }
    \label{fig:unprojection_diagram}
\end{figure}
\section{Related Work}
\label{sec:related_work}
We divide our discussion of related work into two main topics -- visualization of high-dimensional data (Sec.~\ref{sec:related_work:high_dims}) and deep learning latent spaces (Sec.~\ref{sec:related_work:nn_latent_spaces}).
\subsection{Visualization of High-Dimensional Data}
\label{sec:related_work:high_dims}
We first list the notation used for the remainder of the paper. Let $\mathbf{x} \in \mathbb{R}^d$ be a $d$-dimensional data point, also called a sample or an observation. Let $D = \{\mathbf{x}_i\}$, $1 \leq i \leq N$ be a dataset of $N$ such samples. The need to examine, interpret, and explore high-dimensional datasets is not new. As early as the 1970's, Andrews\cite{andrews:1972:plot_greater_than_2d} recognized the need to visualize data whose dimension $d$ exceeded the limits of what can conventionally be drawn on a 2D plane. Geng\cite{geng:2013:3d_display} presented several techniques and systems from stereoscopic imaging to volumetric displays that allow visualizing $d=3$ dimensional data. Yet, as the dimensionality $d$ grows beyond three or four, it becomes clear that increasing the dimensionality of display technology is not a solution.

\subsubsection{Projections}
\label{sec:related_work:projection}
Also called Dimensionality Reduction (DR) methods, projections are techniques that aim to go beyond the aforementioned limitations of high-dimensional visualization techniques\cite{gorban:2008:principal_manifold_techniques,van:2009:dim_reduction_survey,joia:2011:lamp, silva:2012:user_centered_projection, sorzano:2014:dim_reduction_survey, sacha:2016:dim_reduction_interaction_survey, jeong:2009:ipca}. Formally put, a projection technique $P : D \rightarrow \mathbb{R}^m$ is a function that maps every point $\mathbf{x} \in D$ of a high-dimensional dataset to a low-dimensional counterpart $P(\mathbf{x})$. Typically $m \in \{2,3\}$, which allows directly depicting the projection $P(D) = \{ P(\mathbf{x}) | \mathbf{x} \in D \}$ as a 2D or 3D scatterplot, respectively. 

Projection techniques $P$ aim to preserve the so-called \emph{data structure} between the original dataset $D$ and its low-dimensional counterpart $P(D)$. Structure is captured in terms of inter-point distances\cite{roweis:2000:lle,paulovich:2008:least,joia:2011:lamp}, point neighborhoods\cite{maaten:2008:tsne,mcinnes:2018:umap}, or clusters\cite{paulovich2006text}. Projections can be further classified as linear\cite{cunningham:2015:linear_dim_reduction_survey} \emph{or} nonlinear\cite{yin07_survey,van:2009:dim_reduction_survey}. Linear techniques, such as PCA, are simple and fast to compute, have an intuitive geometric interpretation, and robust association with statistical analysis. Nonlinear techniques, such as UMAP, are generally more computationally expensive, but strive to represent local neighborhood information with minimal distortion. There are also a number of projection techniques that generally fit under the RadViz family\cite{hoffman:1997:radviz, angelini:2019:enhancing_radviz, pagliosa:2019:radviz++}. RadViz is able to visualize multidimensional data in 2D by anchoring each feature around the perimeter of a circle, and leverages spring forces from those points to assign each instance a location inside the circle. Projection techniques are further classified, analyzed, and compared both theoretically and practically in a number of surveys\cite{hoffman02,bunte11,sorzano:2014:dim_reduction_survey,maljovec15,nonato18,espadoto19}, to which we refer the reader.

All projection techniques $P$ transform data between the original space $D$ and the projection space $\mathbb{R}^m$. Several techniques aim to show errors in this process, \emph{i.e.}, areas in $P(D)$ that may miss or not reflect actual structures in $D$. For example, Stress Maps\cite{seifert:2010:stress_maps} is a visual analysis tool that displays the local stress values, or how local distance relationships have changed, under a projection algorithm. Other error metrics and subsequent visualization mechanisms include trustworthiness and continuity\cite{venna10}, false and missing neighbors\cite{martins14,martins15}, and false neighborhoods and tears\cite{aupetit:2007:visualizing_distortions_in_projections,lespinats:2011:checkviz}. Surveys of such metrics are given in\cite{nonato18,espadoto19}. In order to demonstrate potential changes of $P(D)$ caused by a hypothetical perturbation of the data in $D$, DimReader\cite{faust:2018:dim_reader} utilized a filled contour plot in the background. 
t-viSNE\cite{kerren:2020:tvisne} focuses on helping users understand t-SNE projections, such as how hyperparameters affect the properties of the final projection. Probing Projections\cite{stahnke:2015:probing_projections} allows users to display the value of any attribute through a background heat map, and also enables users to correct distance errors in the projection by moving individual points in $P(D)$ on the 2D projection space. A similar technique is proposed by LAMP\cite{joia:2011:lamp}. In contrast, Dis-Function\cite{brown:2012:dis} updates the mapping $P$ from the user's dragging of data points to generate new, and hopefully better, projections.
Sirius\cite{dowling:2019:sirius} allows practitioners to investigate both the observations and attributes of a dataset through symmetric projections.

Choosing a good projection -- one which yields a low projection error on a given family of datasets, is simple to use in terms of parameter setting, is robust to small changes in the data $D$, and is computationally scalable to large dimensions $d$ and sample counts $N$ -- is challenging. Recently, Neural Network Projection (NNP)\cite{nnp} was proposed as a method to achieve these goals by leveraging deep learning: Given a dataset $D$, a small subset $D_S \in D$ is chosen and projected by any user-chosen technique $P$. After a suitable projection $P(D_S)$ is obtained by tuning $P$'s parameters, a fully-connected feed-forward neural network is trained to infer $P(D_S)$ from $D_S$. The trained network is then used to project any data drawn from a similar distribution as $D_S$. NNP has shown remarkable ability in producing projections that mimic a wide range of techniques $P$ on many types of datasets, with little or no parameter tuning\cite{espadoto:2020:innp}. Moreover, NNP is parametric, making it robust to small-scale data changes in $D$ while also providing an out-of-sample capability -- that is, NNP learns a \textit{continuous} function $P : \mathbb{R}^n \rightarrow \mathbb{R}^m$ with $n << m$ rather than a discrete mapping formed by a non-parametric projection. NNP is important as a basis for the discussion of inverse-projections in the next section.

\subsubsection{Inverse Projection}
\label{sec:related_work:back_projection}
Inverse-projection can be seen as a function $P^{-1} : \mathbb{R}^m \rightarrow \mathbb{R}^n$, which should ideally be the mathematical inverse of a given projection $P$, \emph{i.e.}, $P^{-1}(P(\mathbf{x})) = \mathbf{x}, \forall \mathbf{x} \in D$. 
A crucial component of inverse-projections is that they \emph{should} have an out-of-sample ability that can be expressed as a continuous mapping, which is generally not the case for direct (discrete) projections. 
Thus, $P^{-1}$ can be used to invert points that fall \emph{between} the points of the scatterplot $P(D)$, helping the user to understand what kind of data samples could project at a particular location in $P(D)$. This ability further supports applications such as data augmentation and classifier exploration\cite{rodrigues:2018:classifier_boundaries,rodrigues:2019:classifier_boundaries}.

Inverse-projection is inherently harder than direct projection due to the need for an out-of-sample ability and the fact that $P^{-1}$ needs to synthesize a high number of dimensions $d$ from a lower dimension  $m$.
Early on, autoencoders were proposed to jointly infer both $P$ and $P^{-1}$ by deep learning to minimize the projection error from $\mathbb{R}^n$ to $\mathbb{R}^m$\cite{hinton2006reducing}. 
While autoencoders are parametric, the resulting mappings are not always intuitive\cite{van:2009:dim_reduction_survey} and autoencoders can be difficult to train\cite{vernier20}. Amorim \emph{et al.} approach inverse-projection in iLAMP\cite{amorim:2012:ilamp} by using local affine transformations, following the earlier idea of the LAMP direct projection technique\cite{joia:2011:lamp}. Mamani \emph{et al.} also use local affine transformations in their inverse-projections as a part of their work on user-driven feature space transformation\cite{mamani:2013:user_driven_feature_space_transformation}. iLAMP was later extended by leveraging radial basis functions (RBFs) to provide a smoother inverse mapping $P^{-1}$, which was shown to be useful for data augmentation\cite{amorim:2015:rbf}. Kriegeskorte and Mur\cite{kriegeskorte:2012:inverse_mds} proposed inverse MDS, which infers pairwise dissimilarities from multiple 2D arrangements of items.
Cavallo \emph{et al.}\cite{cavallo:2018:praxis} used inverse-projection in Praxis, an interactive exploratory analysis tool for high-dimensional data. The authors leveraged the analytical inverse of PCA in addition to an autoencoder to both project and inverse-project data.
Similarly, Zhao \emph{et al.}\cite{zhao:2020:chartseer} used a Grammar Variational Autoencoder (GVAE)\cite{kusner2017grammar} to project and inverse-project data charts for steering exploratory visual analysis.

%
\subsection{Latent Spaces with Neural Networks}
\label{sec:related_work:nn_latent_spaces}
Recent developments in machine learning and AI have shown that deep learning approaches are both accurate and flexible when used and trained properly\cite{goodfellow19}.
In general, neural network encoders work by learning a mapping from the input data space $\mathbb{R}^d$ to a lower dimensional representation $\mathbb{R}^m$ called the \emph{latent space}.
This mapping, conceptually similar to our projection $P$, is often difficult to interpret as the latent dimensions are abstract. Moreover, the neural network's operation is harder to understand than the equivalent operation of a typical projection function $P$. 
\subsubsection{Interpreting the Latent Space}
\label{sec:related_work:interpreting_latent_space}
Interactive visualization tools have been developed to help with analysis tasks that give a better understanding of latent spaces. In particular, when a neural network model has a generative component (\emph{e.g.} autoencoders and Generative Adversarial Networks), its latent space can be explained by bringing data points back to the original space via its generative component.
Liu \emph{et al.}\cite{liu:2019:latent_space_catography} presented a latent space cartography (LSC) visual analysis system for vector space embeddings.
The LSC system was created to address common interpretation tasks for latent spaces. It provides a means to both quantify attribute vector uncertainty and compare multiple attribute vectors.
Spinner \emph{et al.}\cite{spinner:2018:towards_interpretable_latent_space} also used latent spaces to visually compare autoencoders with variational autoencoders. A number of techniques have been developed in order to try to disentangle the latent features of autoencoders\cite{higgins:2017:beta_vae,kim:2019:disentangling_factorising,chen:2019:isolating_disentanglement_vae}. A recent work by Gou \emph{et al.} moved these advances forward within a full visual analytics system for traffic light detection\cite{Gou:2020:valtd}. Additional visualizations making use of, and explaining, latent spaces are discussed in a recent survey\cite{garcia18}.

\section{Learning the Inverse Projection}
\label{sec:method}

\begin{figure*}[t!]
    \centering
    \includegraphics[width=0.9\linewidth]{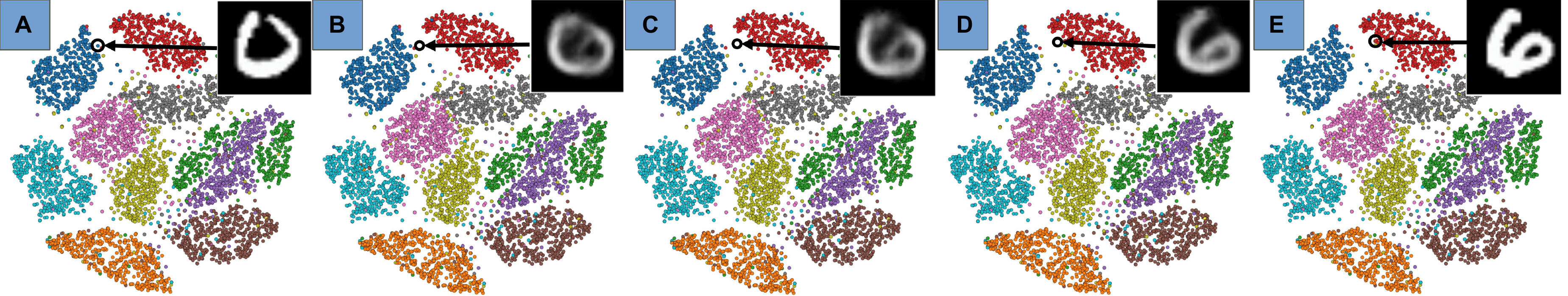}
    \caption
    {
        Example of back projection-enabled interpolation in the original space, as a user explores regions between original data points.
        %
        As the user moves the mouse from a  point representing the digit 0 (A) to a point representing the digit 6 (E), the pixel under the mouse is used as input to the back projection.
        B-D show how, as the user moves closer to the original point, the recovered high-dimensional data is meaningfully interpolated.
    }
    \label{fig:interpolation_table}
\end{figure*}
%

%
%
Figure~\ref{fig:unprojection_diagram} shows  the operation of \sys. Given a dataset $D \subset \mathbb{R}^d$, of $N$ points, let $P(D) = \{ P(\mathbf{x}_i) | \mathbf{x}_i \in D\}$, $1 \leq i \leq N$ be its projection by any user-chosen projection method $P$. In practice, $P(D)$ is a $m=2$ dimensional scatterplot, so $P(D) \in \mathbb{R}^2$. \sys constructs an approximation $B : \mathbb{R}^2 \rightarrow \mathbb{R}^d$ of the inverse $P^{-1}$ of $P$ by using deep learning. Let
\begin{align}
    \hat{\mathbf{x}} = B(\bm{\theta}, \mathbf{y})
\end{align}
be a $d$-dimensional point $\mathbf{\hat{x}}$ inferred by the neural network $B$ from a 2D point $\mathbf{y}$. Here, $\bm{\theta}$ are the learned parameters of the function $B$ (\emph{i.e.}, the weights of the network). To train the model, we minimize the loss between each predicted $\hat{\mathbf{x}}_{i}$ and true $\mathbf{x}_{i}$ within the training set ($D_s \subset D$, $P_s \subset P(D)$) using some loss function.
%
%
%
\subsection{Data}
\label{sec:method:data}
We used five different datasets across our evaluation and proposed applications.\\

\vspace{-0.2cm}
\noindent\textbf{MNIST}: This dataset\cite{lecun:2010:mnist} has $N=70K$ grayscale images of hand-drawn digits, zero through nine. Each image is at a resolution of $28 \times 28$. The images have been translated so that the center of mass of the pixels is at the center of the image. The MNIST dataset is commonly used to illustrate and measure the quality of projection techniques\cite{maaten:2008:tsne,van:2009:dim_reduction_survey,espadoto19,nnp,espadoto:2019:nn_inv}.\\

\vspace{-0.2cm}
\noindent\textbf{Fashion-MNIST:} This dataset\cite{xiao:2017:fashion_mnist} is constructed in the same manner as the original MNIST dataset, but contains pictures of different items of clothing. It was designed as a slightly more difficult replacement for the MNIST dataset.\\

\vspace{-0.2cm}
\noindent\textbf{Blobs:} This synthetic dataset has $N=70K$ points sampled from a Gaussian distribution with 5 different centers (clusters) in $d=50$ dimensions.\\

\vspace{-0.2cm}
\noindent\textbf{Sphere:} This dataset consists of $N=8K$ points uniformly sampled from a 3D unit sphere. It allows us to clearly demonstrate the behaviour of the projection techniques included, and more importantly, offer a simple illustration of our gradient map visualization.\\

\vspace{-0.2cm}

\noindent\textbf{Swiss Roll:} This dataset consists of $N=70K$ points sampled from a densely-sampled 2D patch which was smoothly mapped to a ``roll'' in 3D. It is commonly used to gauge the capability of projections to ``unroll'' the data back to its 2D configuration\cite{amorim:2012:ilamp,joia:2011:lamp,balasubramanian2002isomap}.

\subsection{Implementation}
\label{sec:method:practicalities}
%
\begin{table}[htb]
\scriptsize
\centering
\begin{tabular}{|l|l|l|l|l|l|l|}
\hline
\textbf{Dataset}                  & $L_1$ & $L_2$ & $L_3$ & $L_4$ & MAE      & STD      \\ \hline
\multirow{5}{*}{Blobs}    & 64        & 128       & 256         & 512        & 0.036941 & 3e-06    \\
                         & 128       & 256       & 512         & 1024       & 0.036944 & 2.5e-05  \\
                         & 256       & 512       & 1024        & 2048       & 0.036945 & 2.1e-05  \\
                         & 640       & 1280      & 1280        & 640        & 0.03695  & 3.3e-05  \\
                         & 240       & 240       & 240         & 240        & 0.036961 & 3.1e-05  \\ \hline
\multirow{5}{*}{MNIST}  & 128       & 256       & 512         & 1024       & 0.06241  & 0.000425 \\
                         & 640       & 320       & 320         & 640        & 0.062606 & 0.000113 \\
                         & 640       & 320       & 320         & 640        & 0.062787 & 0.000551 \\
                         & 480       & 480       & 480         & 480        & 0.06303  & 5e-05    \\
                         & 480       & 480       & 480         & 480        & 0.063168 & 0.000258 \\ \hline
\multirow{5}{*}{FashionMNIST} & 1024      & 2048      & 4096        & 8192       & 0.072804 & 0.000411 \\
                         & 1280      & 2560      & 2560        & 1280       & 0.072873 & 0.000136 \\
                         & 512       & 1024      & 2048        & 4096       & 0.073108 & 0.000268 \\
                         & 1280      & 640       & 640         & 1280       & 0.073209 & 6.5e-05  \\
                         & 256       & 512       & 1024        & 2048       & 0.073214 & 0.00064  \\ \hline
\multirow{5}{*}{Swiss}   & 64        & 128       & 256         & 512        & 0.011698 & 0.000489 \\
                         & 256       & 512       & 1024        & 2048       & 0.012288 & 0.001286 \\
                         & 640       & 1280      & 1280        & 640        & 0.013136 & 0.000805 \\
                         & 160       & 320       & 320         & 160        & 0.013209 & 0.001    \\
                         & 640       & 320       & 320         & 640        & 0.013929 & 0.001523 \\ \hline
\end{tabular}
\caption{Top five configurations per dataset sorted by lowest mean absolute error (MAE), computed by averaging three different runs. Columns $L_i$ show the number of neurons used in the respective hidden layers.}
\label{table:top_architectures}
\end{table}
We next describe the design and tuning of the neural network used to learn the inverse projection. Following\,\cite{elsken:2019:nas_survey}, and also the method used to tune NNP\,\cite{espadoto:2020:innp}, we used grid search to explore different architecture configurations: total number of neurons,  neurons per layer, and dropout values.
We ran the grid search across the four datasets introduced in Sec.~\ref{sec:method:data}. As the direct projection $P$, we used t-SNE, which was earlier shown to be the hardest projection from a set of nine different projections to mimic via deep learning\cite{nnp}. Hence, we believe that t-SNE is also a hard challenge to invert via \sys. We varied the training-set size $|D_s|$ between 5250, 10500, 21000, and 42000 samples. To account for variation in random initialization of the neural network weights, we ran each configuration three times and averaged the results into a single error score.
We measure quality via mean absolute error (MAE) $\frac{1}{N}\sum_{i=1}^{N} | x_{i} - \hat{x}_{i} |$, and also provide its standard deviation across the three runs. Training is stopped automatically on convergence, defined as the moment when the validation loss stops decreasing. We next discuss the hyperparameters investigated.\\

\noindent\textbf{Network Architectures}: We restricted ourselves to fully-connected layers and used four hidden layers ($L_1, \ldots, L_4$) in each configuration.
We varied the network shape and number of neurons in each layer. The total number of neurons $\nu$ in each network varied between 240, 480, 960, 1920, 3840, 7680, and 15360.
We experimented with four network shapes (see Table~\ref{tab:shapes}).

\begin{table}[htb]
\scriptsize
\centering
\begin{tabular}{ | l | l | l | l | l |}
\hline
\textbf{Shape} & $ |L_1|$ & $ |L_2|$ & $ |L_3|$ & $ |L_4|$ \\
\hline
straight & $\nu/4$ & $\nu/4$ & $\nu/4$ & $\nu/4$ \\
wide & $\nu/6$ & $\nu/3$ & $\nu/3$ & $\nu/6$ \\
bottleneck & $\nu/3$ & $\nu/6$ & $\nu/6$ & $\nu/3$ \\
fan-out & $\nu/15$ & $\nu/7.5$ & $\nu/3.84$ & $\nu/1.875$ \\
\hline
\end{tabular}
\caption{Shapes of the tested networks for learning NNInv.}
\label{tab:shapes}
\end{table}

\noindent\textbf{Activation Functions}: We used a ReLU activation function for all hidden layers. Since the input data $D$ is normalized such that each of the $d$ dimensions ranges over $[0,1]$, we used a sigmoid activation function on the output layer.\\
\noindent\textbf{Regularization}: We used both early stopping and dropout, with dropout probabilities of $0.125$, $0.25$, $0.5$. Experiments showed dropout was not generally effective. We believe that this is due to the fact that overfitting is unlikely given that we used smaller networks and early stopping.\\ 
\noindent\textbf{Loss Function}: For the loss function we used MAE. \\
\noindent\textbf{Optimizer}: We used the Adam optimizer\cite{kingma_adam_2014}, given its good performance with NNP\cite{nnp,espadoto:2020:innp}.\\
Table~\ref{table:top_architectures} shows the MAE and standard deviation (STD) results for the top-five configurations of the $1536$ tested ones, \emph{i.e.}, the ones obtaining lowest error. The full results including all configurations tested is available in the supplemental material.
The best architectures for each dataset either had the same number of neurons in each layer, or used a widening architecture which doubles or quadruples the number of neurons in each successive layer.

Our results suggest that smaller architectures can be used other than the original architecture from Espadoto \emph{et al.}\cite{espadoto:2019:nn_inv}.
The only dataset that performs better with more than 1920 neurons is the FashionMNIST dataset. For this dataset, we obtain a slightly lower MAE when using 7680 neurons. However, as in most cases observed, the error decrease is negligible compared to the increase in complexity (network size).
%
Summarizing our findings from the tested datasets, we offer suggestions for future experimentation: (1) networks should follow a straight or fan-out style shape as described in Table~\ref{tab:shapes}, and (2) even relatively small networks can perform exceptionally well at this task as seen in Table~\ref{table:top_architectures}.

\section{Applications of Inverse Projection in Visual Analytics}
\label{sec:application}
Traditional error calculations may tell something of the overall loss, \emph{i.e.}, going from high-dimensional space to 2D and back to the original space. 
However, a robust analysis of an inverse-projection technique must include more than just this type of error. 
This section focuses on a qualitative evaluation of inverse projections using applications that are of interest to the visualization community.
In particular, we explore three use cases of inverse-projection in visual analytics: (1) direct interpolation of high-dimensional data using the 2D screen, (2) leveraging the generation of high-dimensional data across the screen to per-pixel color classifier agreement, and (3) using the generated high-dimensional data to illustrate high gradient areas of the projection.
\subsection{Case Study 1: Dynamic Imputation}
\label{sec:application:reconstruction}
One shortcoming of the current use of projection methods is that the projections are ``one-way streets.''
From a user interaction and exploration standpoint, the most that a user can do using such techniques is to select a data point in the (2D) visualization and look up the original values of that point in high-dimensional space.
Due to this limitation, the user's exploration of the data is restricted. 
For example, the user would have no easy way of knowing why two data points appear close to each other in the 2D space, or what other data points, if they existed, would appear near or between these points.
\subsubsection{Example with MNIST}
In this case study, we demonstrate the use of inverse-projection to perform ``dynamic imputation.'' 
The inspiration for this case study comes from recent works by Cavallo \emph{et al.}\cite{cavallo:2018:praxis} that explores inverse-projection with PCA and autoencoders, and Kwon \emph{et al.}\cite{kwon:2020:deep_generative_graphs} that generates graph layouts from the user's interactions in a 2D latent space.

Consider Figure~\ref{fig:interpolation_table}: The user can select projected data points in a 2D visualization (of the MNIST dataset) and see their original values (see Figure~\ref{fig:interpolation_table}A and Figure~\ref{fig:interpolation_table}E), similar to traditional visual analytics systems.
However, with the use of inverse-projection, the user can \emph{also} select an \textit{``empty''} space between these data points (see the three inner images).
The inverse-projection function implicitly performs imputation (\emph{i.e.}, generating a new data point) when performing inference over the 2D pixel location to find its position in high-dimensional space.

\begin{figure}[b]
    \centering     
    \includegraphics[width=0.9\linewidth]{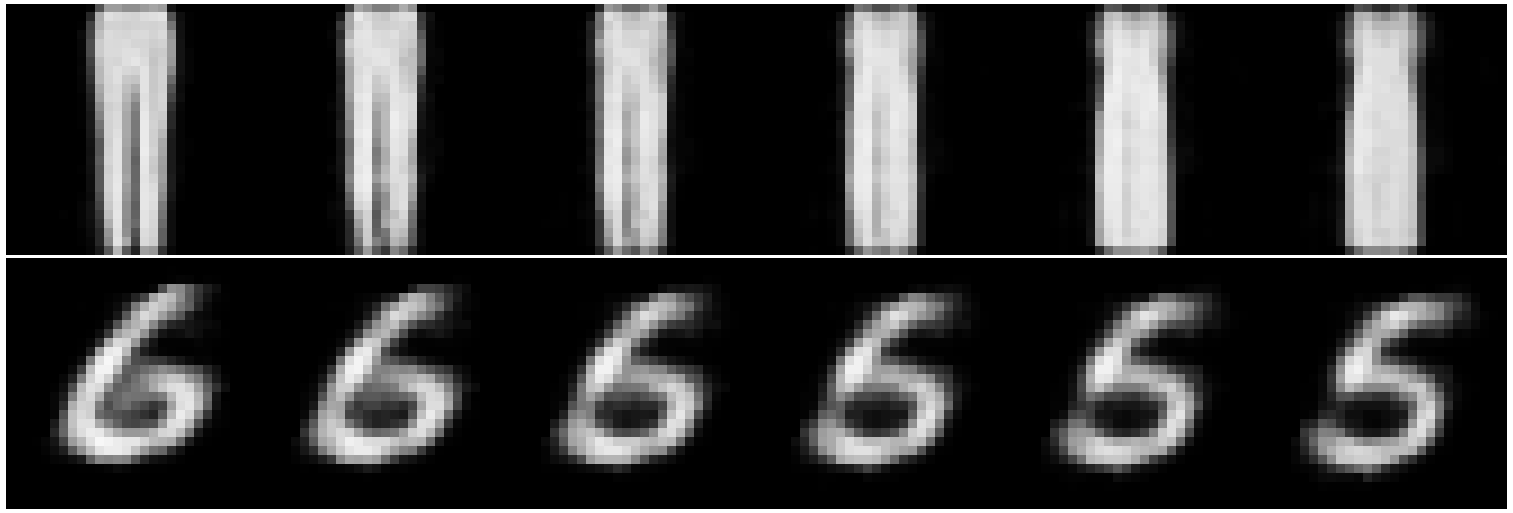}
    \caption{
        Images generated when moving from one cluster to another within a projection, and feeding the x and y coordinates into \sys. Top row: FashionMNIST, moving from class ``pants'' to class ``dress''; bottom row: MNIST, moving from class 6 to class 5.
    }
    \label{fig:interpolation_between_clusters}
\end{figure}

Since the inference step of a trained neural network is fast, this computation can be done in a web browser and be made fully interactive using mouse hovering. In this example, the computation time of inverse-projecting a point in tensorflow.js on an Intel i7-8650U CPU is below 10 milliseconds.
With this high degree of interactivity, the user can quickly explore both the high-dimensional dataset as well as the high-dimensional \emph{space} (between data points) itself.
Figure~\ref{fig:interpolation_between_clusters} further showcases the ability of using inverse-projection to interpolate between clusters. Here, the images furthest to the left and right represent two visually distinct objects (\emph{i.e.}, a pair of pants and a dress, and the digits 6 and 5).
The images in between are interpolations generated by the inverse-projection algorithm.

\begin{figure}[t]
    \centering
    \includegraphics[width=0.8\linewidth]{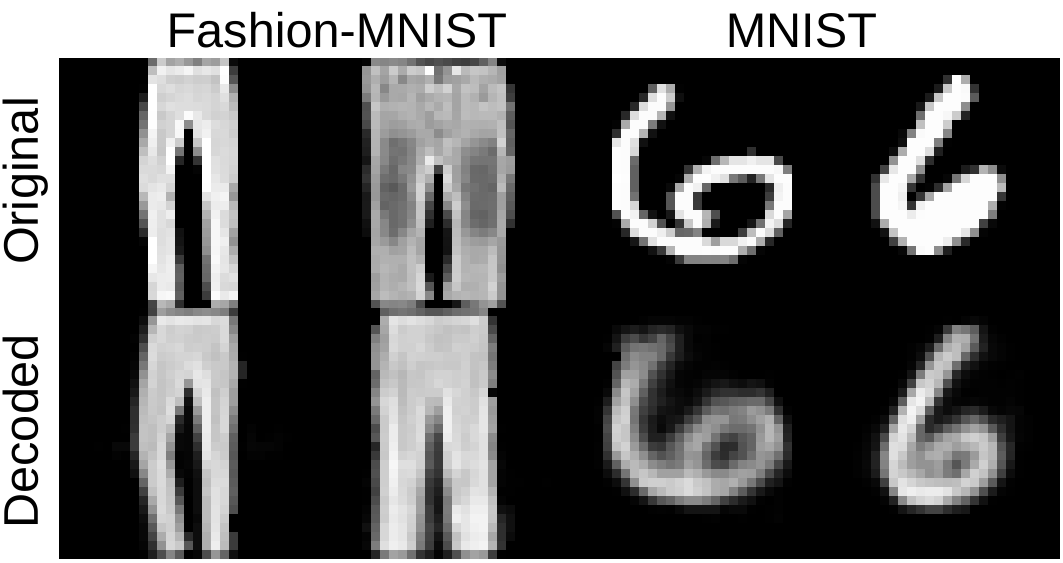}
    \caption
    {
        Data points from the test portion of both datasets. The images labeled as ``Generated'' illustrate the inverse-projection corresponding to the projection of the images labeled as ``original''.
    }
    \label{fig:interpolation_across_clusters}
\end{figure}

\subsubsection{Evaluation of the Inverse Projection}
Using this framework, we can also visually evaluate the quality of the inverse-projection algorithm.
Specifically, instead of selecting an ``empty'' pixel, a user can select the 2D position of an existing data point.
We can then compare the original values of the data point with the values generated by the inverse-projection algorithm.
For example, the two images on the upper left side of Figure~\ref{fig:interpolation_across_clusters} are two different styles of pants from the original Fashion-MNIST dataset.
The two images directly below, on the lower left side, are those generated by the inverse-projection. Similarly, images on the upper right side of Figure~\ref{fig:interpolation_across_clusters} are from the original MNIST dataset, and images on the lower right side of Figure~\ref{fig:interpolation_across_clusters} are generated. In both cases, the generated images are ``blurrier'' than the originals. 
However, it is shown that the inverse-projection function has successfully learned the important visual features of these images and can reproduce them with high fidelity.
\subsubsection{Implications to Visual Analytics}
Although we used two relatively simple image datasets in this case study (MNIST and Fashion-MNIST) for illustrative purposes, the use of inverse-projection for dynamic imputation should be extendable to visual analysis of other high-dimensional datasets, including temporal data, geographic data, and tabular data.
As such, having an accurate inverse-projection function in a visual analytics system can allow the system designer and the user to explore high-dimensional data in ways that have not been possible.
For example, in the context of business analysis, the use of inverse-projection for data imputation can serve as a \textit{``hypothesis generator''} (\emph{e.g.}, Figure~\ref{fig:interpolation_table}). 
With inverse-projection, the user can interpolate between the 0 and the 6 from the original data, and use inverse-projection to generate hybrid examples between.
While the generated data points are estimates of the inverse-projection function, they may nonetheless serve as potential hypotheses for an analyst to further explore.
\subsection{Case Study 2: Model Agreements}
\label{sec:application:model_agreements}

\begin{figure*}[t]
\centering
\includegraphics[width=0.85\linewidth]{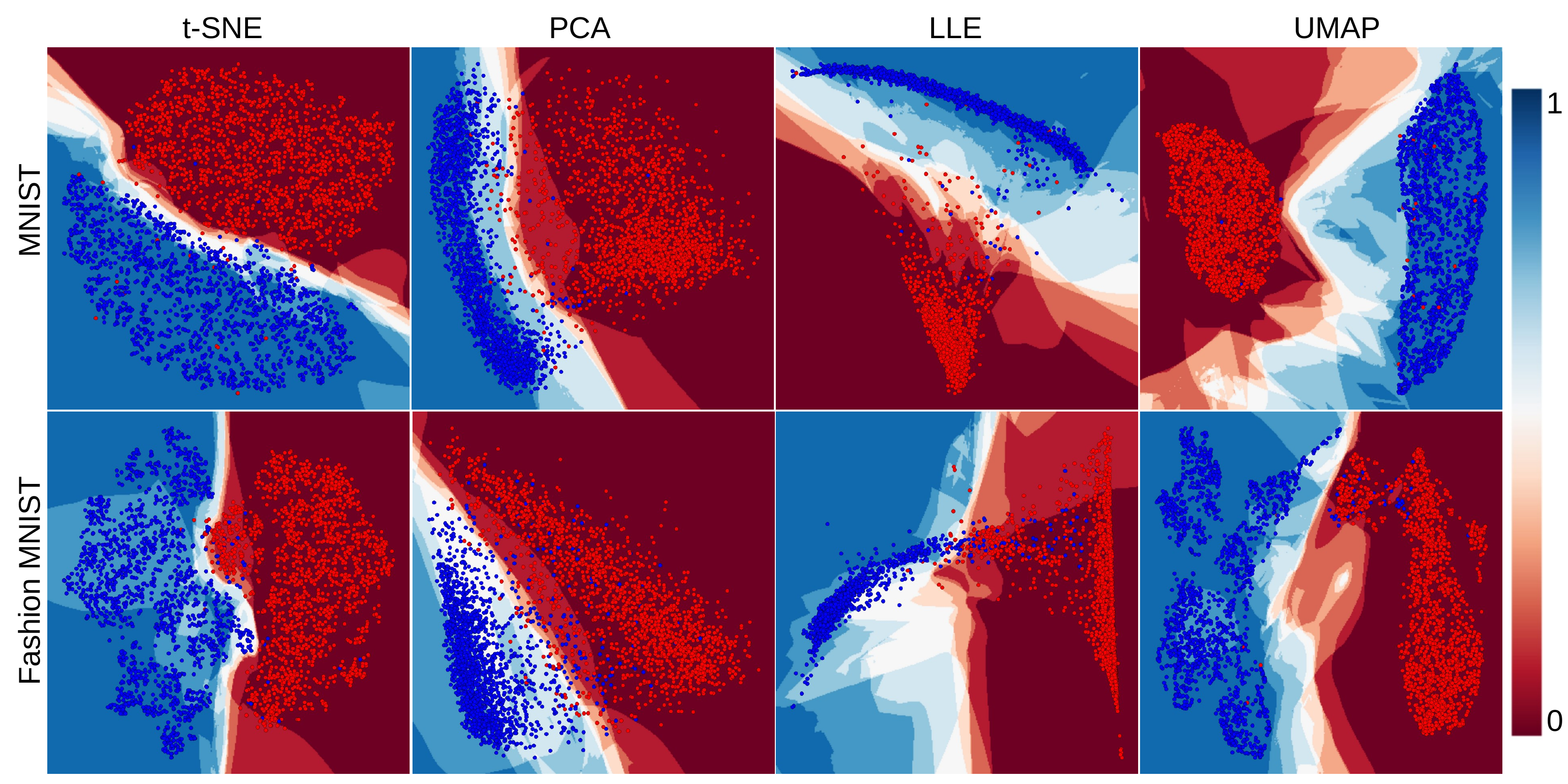}
\caption{
Classifier agreement map.
Top row shows the result of classification of two digits in the MNIST data (digits 1 and 7), bottom shows classification of two objects in the  MNIST-Fashion data (handbags and shirts).
Color in these images denote agreement between 9 classifiers.
Red represents agreements for class 1 and blue for class 2.
More saturated colors indicate higher agreement. 
In between clusters where agreement is low, the colors tend to be desaturated (white). 
}
\label{fig:classifier_table}
\end{figure*}

Previous work in defining and interpreting back projections has shown that creating dense pixel maps in the 2D projection space can provide additional insight into the behavior of classification type tasks\cite{espadoto:2019:nn_inv, rodrigues:2018:classifier_boundaries, rodrigues:2019:classifier_boundaries}. Figure~\ref{fig:classifier_table} shows how this concept is extended to highlight regions of lower classification agreement.
\subsubsection{Example with MNIST and Fashion-MNIST}
We demonstrate ensemble classification confidence by creating dense pixel maps to show the classifier agreement of two of the ten digits in the MNIST dataset (digits 1 and 7) and two of the objects in the Fashion-MNIST dataset (handbags and shirts).
While there is nothing preventing the technique from being extended to multiclass classification, as in previous work\cite{espadoto:2019:nn_inv, rodrigues:2018:classifier_boundaries, rodrigues:2019:classifier_boundaries}, we limit ourselves to binary classification.
In both cases, we begin by inverse-projecting each screen pixel to learn its position in the high-dimensional space.
This high-dimensional point is then put through some number (greater than one) of classification methods.
Since our dataset only contains two classes, each of the classifiers will simply assign a data point to one class or the other.
We then color the pixel based on the number of classifiers that predicted each class.
As shown in Figure~\ref{fig:classifier_table}, we color a pixel bright blue if the majority of classifiers predicted class one, and bright red if the majority of classifiers predicted class two.
In between these two extremes, pixels are colored by decreasing the amount of saturation such that complete disagreement between the models results in a white pixel -- that is, half of the classifiers says the data point inverse-projected from the respective pixel belongs to class 1, while the other half says the point belongs to class 2.
The ensemble is formed by nine classifiers, namely Logistic regression, Linear SVM, SVM with radial basis function, K-Nearest Neighbors, Gaussian Process, Decision Tree, Random Forest, Adaboost, Gaussian Naive Bayes, and Quadratic Discriminant Analysis. These classifiers represent a diverse number of classification algorithms, including linear and non-linear methods.
The output from the nine classifiers is used to generate the images in Figure~\ref{fig:classifier_table}, where not only can we see the class memberships of each point, we can also see the shape of the decision boundaries.
\begin{figure}[h]
\centering
    \includegraphics[width=0.55\linewidth]{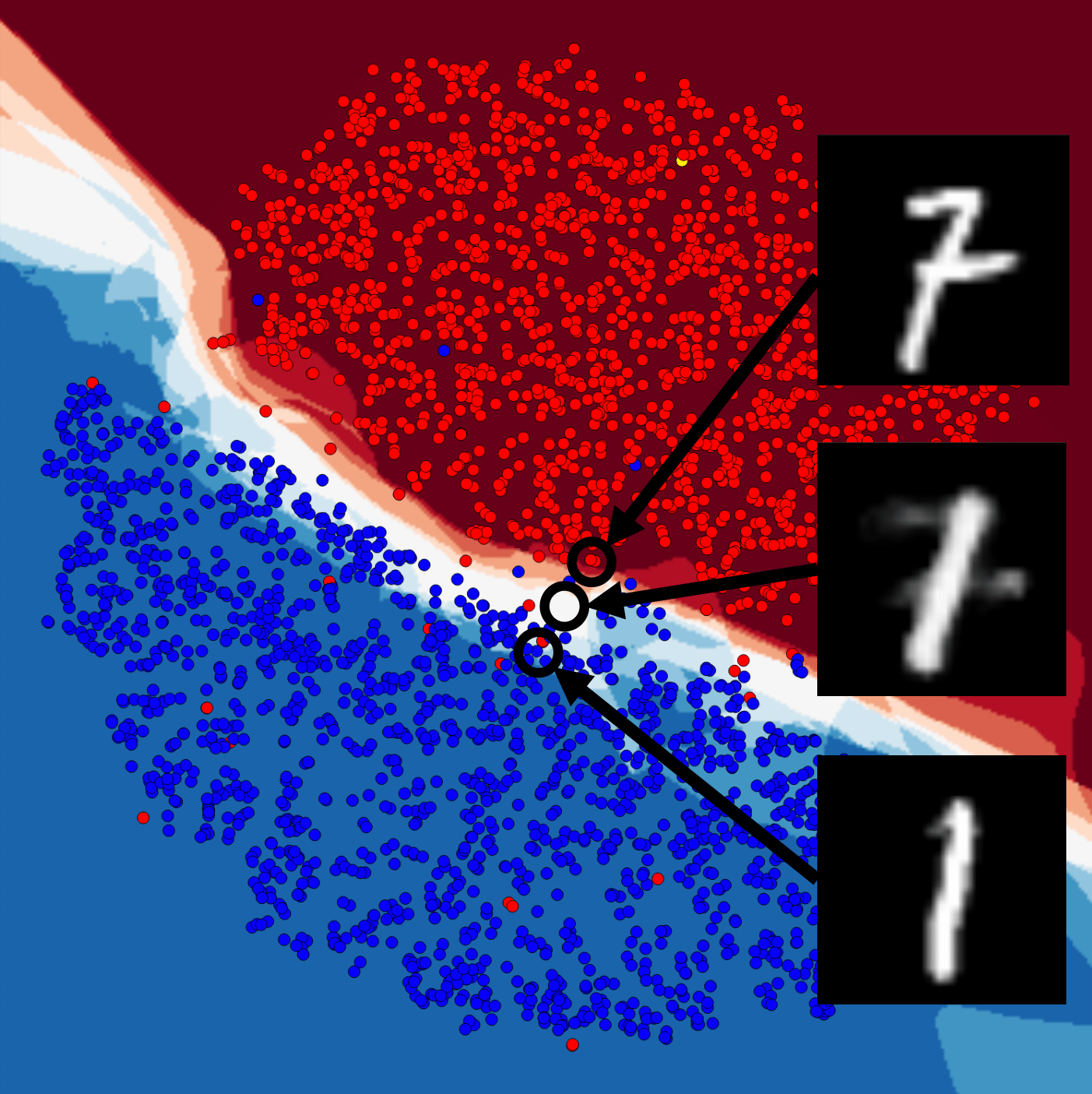}
    \caption
    {
        Visual inspection of points in the decision boundary as yielded by the ensemble classifier and drawn using back projection.
    }
    \label{fig:decision_boundary_closeup}
\end{figure}
For example, we can combine the use of inverse-projection for visualizing decision boundaries with its use for dynamic imputation, resulting in an interactive visual exploring system for understanding the uncertainty of the classifiers.

As an illustrative example, consider the differences between t-SNE and UMAP in Figure~\ref{fig:classifier_table}. 
When only considering the separability of the two clusters, one would likely assume that UMAP outperforms t-SNE, especially for the MNIST dataset (top row of Figure~\ref{fig:classifier_table}).
However, when inspecting the decision boundaries, it becomes less clear that the separability affects the classifiers' abilities to distinguish data points from one class to another.
Specifically, in the t-SNE example, although the separation between the two clusters in the MNIST data is small, the boundaries are sharp and clean.
Conversely, while UMAP produces high separation between the two clusters, there are disagreements between the classifiers in that space.
\subsubsection{Implications to Visual Analytics}
While there has been a number of proposed methods for illustrating the decision boundaries for classifiers of high-dimensional data\cite{migut:2015:decision_boundaries, Hamel:2006:decision_boundaries_of_svm, rodrigues:2018:classifier_boundaries, schulz:2015:decision_boundaries, rodrigues:2019:classifier_boundaries}, our proposed use of inverse-projection offers an alternative that can be more flexible for visual analytics systems.
As illustrated in Figure~\ref{fig:decision_boundary_closeup}, the user can hover over areas with low model agreement (\emph{e.g.}, pixels that are white or near white), and see what characteristics of the data might cause the classifier models to disagree.
In the context of designing new visual analytics techniques, the use of inverse-projection to help users better understand the behaviors of machine learning models can prove to be invaluable.
Colloquially referred to as Explainable AI (or XAI), visualization researchers have been active in developing novel visualization and interaction techniques that can help a user understand, debug, and improve a complex machine learning model.
While the space of XAI is large, we posit that the inverse-projection technique can contribute to this broad space of research. 
\begin{figure}[t]
\centering
    \includegraphics[width=1.0\linewidth]{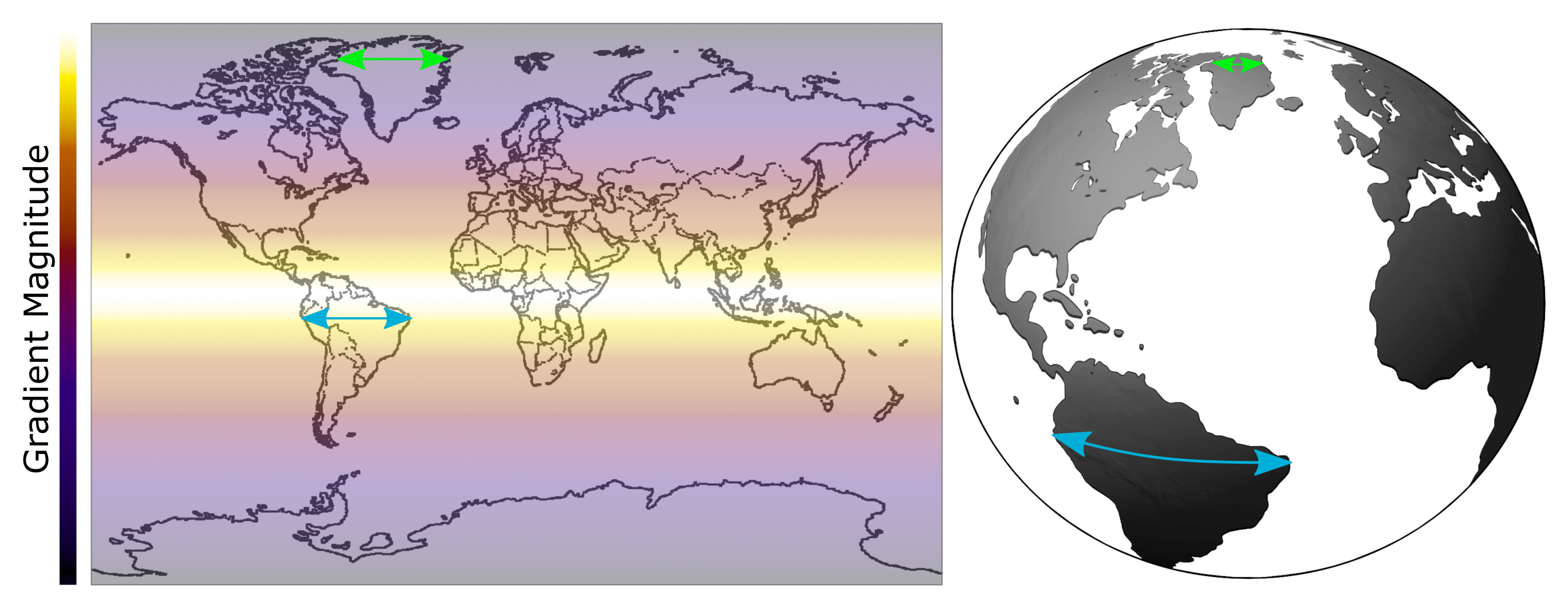}
    \caption
    {
        Two equal length line segments (green and blue) are placed on a Mercator projection of the Earth overlaid with its gradient image.
        After inverse projection, the lengths of these segments are dramatically different.
        The degree of change corresponds to the difference in the pseudo total derivative, shown by the gradient image overlay in the projection.
        Note that the green segment (in a region of low gradient) shrinks \emph{vs} the blue segment (in a high gradient region around the equator).
        This figure is an illustrative example of how gradient images function (note that the figure does not reflect the actual gradients of the sphere).
    }
    \label{fig:globe_mapping}
\end{figure}
\begin{figure}[h]
    \centering
    \includegraphics[width=0.95\linewidth]{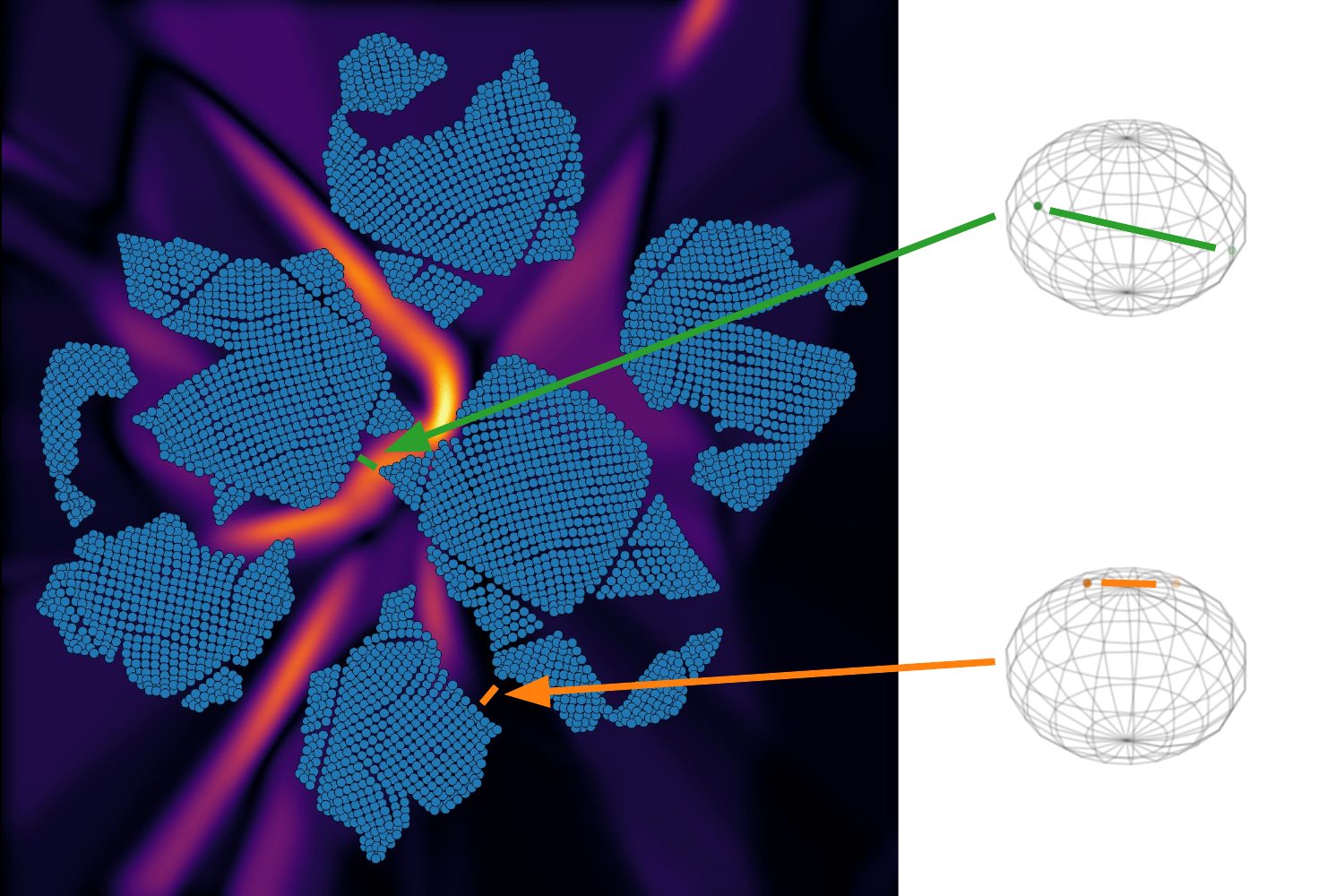}
    \caption
    {An illustration of how illuminated areas in the gradient map show areas where a small change in the low-dimensional space signifies a large change in the high-dimensional space. On the left, there is a 2-dimensional t-SNE projection of a 3-dimensional sphere, with a gradient map underlaid. Two similar distances are highlighted, one that crosses a highlighted area (green line), and one that crosses a dark area (orange line). On the right, two spheres are drawn, one for the green area, and one for the orange area. The dots on the spheres are the two closest points to the green and orange lines in the projection. The points closest to either end of the green line on the projection are much further away on the sphere than the points closest to either end of the orange line.} 
    \label{fig:gradient_distances}
\end{figure}
\subsection{Case Study 3: Gradient Map Visualization}
\label{sec:application:model_gradient}
\begin{figure*}[h]
\centering
\includegraphics[width=0.8\linewidth]{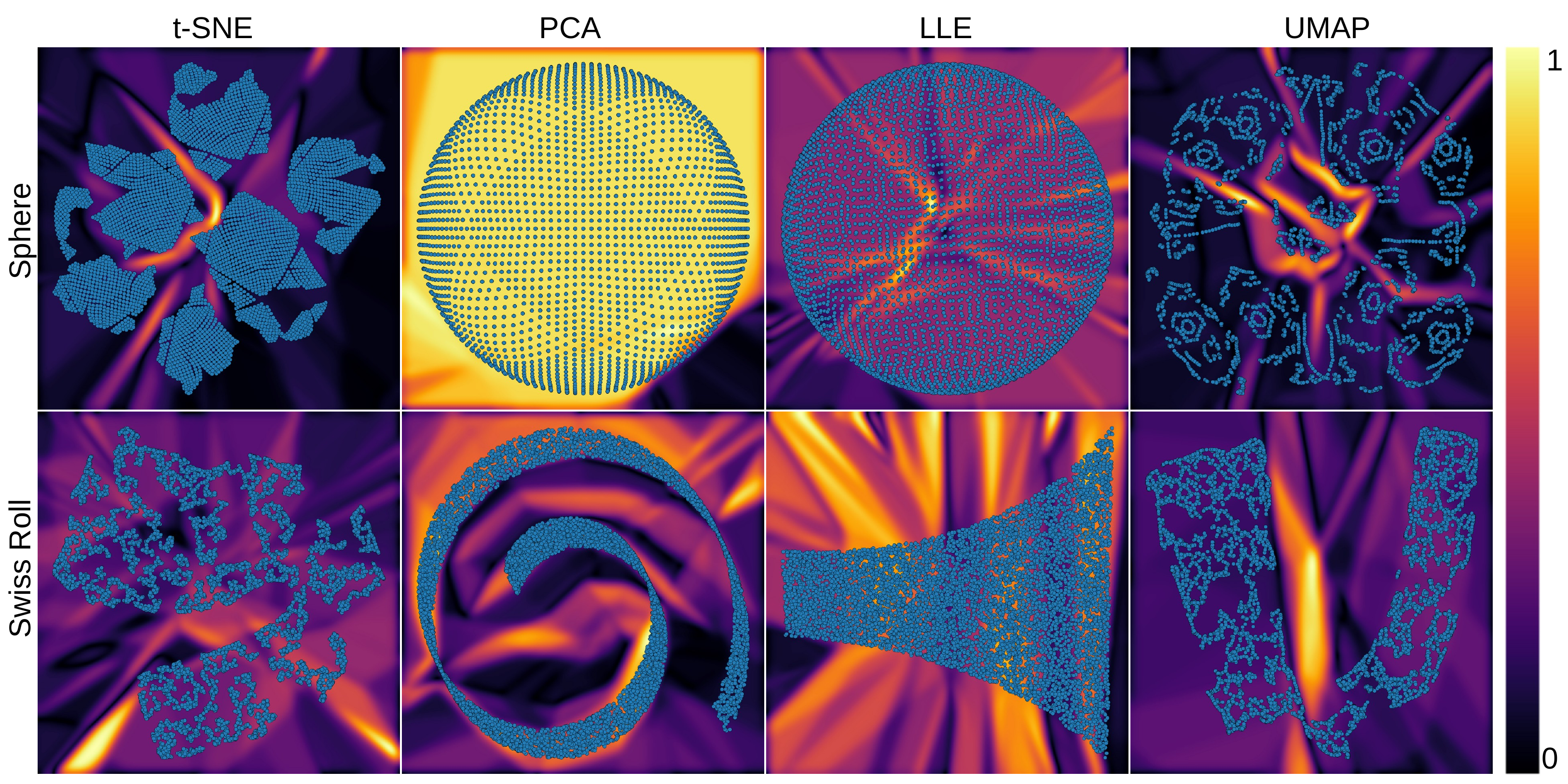}
    \caption{Gradient maps for 3D sphere dataset (top row) and 3D swissroll dataset (bottom row), showing the pseudo total derivative of the neighborhoods of all inverse-projected data points. Blue dots represent the points projected from three dimensions to two. The gradient at each pixel is determined by inverse-projecting each pixel's neighborhood and computing the gradient in the recovered high-dimensional space. Overall, the gradient maps show areas in the projections with high rates of change, suggesting peaks or valleys in the embedding space.}
    \label{fig:gradient_table}
\end{figure*}
Related to Explainable AI (XAI), one of the primary use cases for multidimensional projection is the visualization and interaction of data that exists in high-dimensional spaces that humans have difficulty interpreting.
Unfortunately, a side effect of projection is the loss of information.
To help mitigate the consequences of information loss imposed by projection, most techniques strive to maintain local relationships.
In other words, they seek to preserve the relative distances between neighboring data points in the high-dimensional space in the two-dimensional projection.
Of course, keeping these relationships intact after projecting is not always possible.

Using the concepts of data imputation, a more holistic view of how a projection represents the spatial relationships between data points is presented.
The ability to determine high-dimensional coordinates from a projected point in 2D enables a more complete investigation of the consequences of selecting a given projection technique by inspecting its \emph{gradient image} (see Fig.~\ref{fig:gradient_table}). This image $D$ is a 2D scalar field representing a pseudo total derivative of inverse projection function $B$ computed using central differences as
\[
D_x(\mathbf{y}) = \frac{B(\mathbf{y}+(w,0)) - B(\mathbf{y}-(w,0))}{2w}
\]
\[
D_y(\mathbf{y}) = \frac{B(\mathbf{y}+(0,h)) - B(\mathbf{y}-(0,h))}{2h}
\]
\[
D(\mathbf{y}) = \sqrt{\|D_x(\mathbf{y})\|^2 + \|D_y(\mathbf{y})\|^2}
\]
where $\mathbf{y}$ is a point in the 2D projection space and $w$ and $h$ are a pixel's width and height, respectively.
In summary, regions of a projection with large gradient values illustrate where the high-dimensional distance is changing most rapidly with respect to the low-dimensional distance.
Figure~\ref{fig:gradient_distances} demonstrates how values on either side of large gradient values map to larger distances in the original data space, compared to values on either side of small gradient values.
While the above method uses simple finite differences, any method for computing the gradient magnitude of $B$ is appropriate.
\subsubsection{Example with Sphere Data}
Figure~\ref{fig:globe_mapping} shows how, under a standard projection for parameterization, even a simple three-dimensional sphere is transformed into a stretched and squeezed plane.
Here, two equal length lines are placed on the parameterized plane at different locations. 
When each line segment is inverse-projected to recover the coordinates on the sphere, it is clear that the relative lengths of each segment have changed considerably.
The degree of change is more completely understood when it is observed in the context of the gradient map overlaid onto this plane.
In this case, areas towards the poles of the globe intuitively have a gradient that approaches zero, while the equator will have the highest gradient. 
Thus, a line segment back projected from the two-dimensional plane will necessarily shrink; however, a similar line segment positioned near the equator will grow in length. 

In the top row of Figure~\ref{fig:gradient_table}, a uniformly sampled three-dimensional sphere was projected to a two-dimensional plane using t-SNE, PCA, LLE, and UMAP.  
In these images, the sample points on the sphere are represented by blue dots, while the background is colored with the gradient image.
In the cases of t-SNE, LLE, and UMAP, the  projections maintain similar gradient characteristics with respect to neighboring points.
However, there are some points that project to regions of high gradient.  
These regions are inevitable, as tears in the three-dimensional sphere are required in order to represent it on a plane.
Conversely, PCA is a linear projection method that does not seek to preserve neighborhood information between data points. As a result, the gradient map  under the data points is constant and reflects the planar nature of the projection space.
\subsubsection{Implications to Visual Analytics}
The gradient maps shown in Figure~\ref{fig:gradient_table} illustrate the use of inverse-projection to help users see the quality of the projection.
It is relevant to note that the gradient maps do not show the topology of an embedding space created using a projection function, which is the goal of works like Stress Maps\cite{seifert:2010:stress_maps}. 
Instead, these gradient maps represent the \textit{reconstructed} embedding space by the inverse-projection function.
In some cases the inverse-projection function does not perfectly recover the original embedding space. 
For example, the top row of the PCA column in Figure~\ref{fig:gradient_table} shows the reconstruction of a plane created by PCA in the 3D sphere dataset. %
Notice that the reconstructed surface is not perfectly linear as should be the case of PCA projections.
As such, we consider the gradient map as a debugging mechanism similar to tools in the XAI community for debugging machine learning models.
In particular, the gradient map can help data scientists and visual analytics researchers to better understand the effect of projection and inverse-projection when visualizing high-dimensional data.
For example, the top left image in Figure~\ref{fig:gradient_table} shows the projection of a 3D sphere using t-SNE. 
The intense colors denote sharp discontinuities between the parts of the 3D sphere separated by t-SNE.
This information has illustrative values and can be used to help a user better understand the behaviors of a projection function such as t-SNE.
\section{Evaluation}
\label{sec:evaluation}
We present an empirical evaluation of the inverse projection function described in the previous section. In the following, we split each dataset $D$ into a training set $D_s$ and a test set $D_p$. We train \sys using the pair $(D_s, P_s)$. Within Sections~\ref{sec:evaluation:quality},~\ref{sec:evaluation:invmap}, and~\ref{sec:evaluation:scalability} we restrict $P$ to t-SNE, but also explore PCA, LLE, and UMAP in Section~\ref{sec:evaluation:qualitative}. We next evaluate the quality of \sys using various error metrics computed using $D_p$ and $P_s$. 
We next discuss our method in terms of scalability (Sec.~\ref{sec:evaluation:scalability}), quantitative assessment of quality (Sec.~\ref{sec:evaluation:quality}), qualitative assessment of quality (Sec.~\ref{sec:evaluation:qualitative}), and our novel inverse-projection error map (Sec.~\ref{sec:evaluation:invmap}).
%
%
\subsection{Quantitative Assessment of Quality}
\label{sec:evaluation:quality}
Besides being fast, we want an inverse-projection to be \emph{accurate}.
That is, given some ground truth pair $(\mathbf{x} \in \mathbb{R}^d, \mathbf{y}=P(\mathbf{x}) \in \mathbb{R}^2)$, \emph{unseen} during training, we want $B(\mathbf{y})$ to be as close as possible to $\mathbf{x}$. This follows the same idea as normalized stress metrics used to gauge the quality of projections in the literature\cite{sorzano:2014:dim_reduction_survey, van:2009:dim_reduction_survey,espadoto19} and also classical validation of inference models in machine learning. We measure quality in our case by computing the average inverse-projection mean square error $MSE=\|\mathbf{x}-B(P(\mathbf{x}))\|^2 / |D_p|$ over the test set $D_p$. The closer MSE is to zero, the better $B$ is. While we minimized MAE in our loss function, we report MSE here to enable easier comparison to earlier papers\cite{amorim:2012:ilamp, amorim:2015:rbf}.

\begin{figure}[!htp!]
    \centering
    \includegraphics[width=1.0\linewidth]{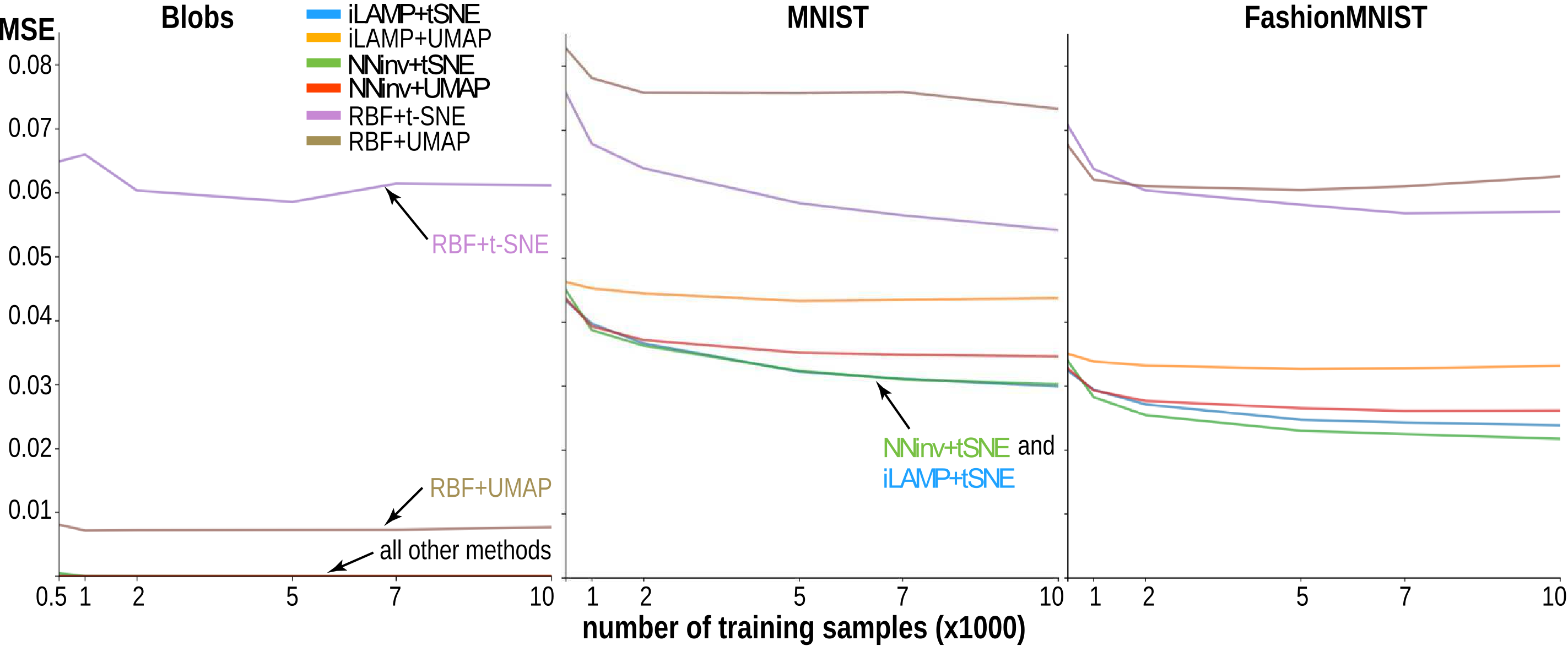}
\vspace{-0.2cm}
    \caption{Mean square error of iLAMP, RBF, and NNInv inverse projections (from\cite{espadoto:2019:nn_inv}). This graphic serves a dual purpose, illustrating how MSE depends on training-set size, and \sys's ability to achieve one of the lowest errors amongst two other inverse-projection methods.}
\vspace{-0.2cm}
    \label{fig:mse_eval}
\end{figure}

Figure~\ref{fig:mse_eval} shows the MSE for our three datasets, two projections (t-SNE and UMAP), and three tested inverse projections (iLAMP, RBF, and \sys). 
We also consider several training-set sizes $|D_s|$ to show how MSE depends on the training amount.
For Blobs, a relatively easy-to-project synthetic dataset, all methods have essentially zero error except RBF.
MNIST and FashionMNIST show similar behavior: Our method achieves consistently one of the lowest errors.
Errors are larger for these real-world complex datasets than for the synthetic Blobs, which is expected.

%
\subsection{Qualitative Exploration}
\label{sec:evaluation:qualitative}
\begin{figure*}[t]
\centering
\includegraphics[width=0.9\linewidth]{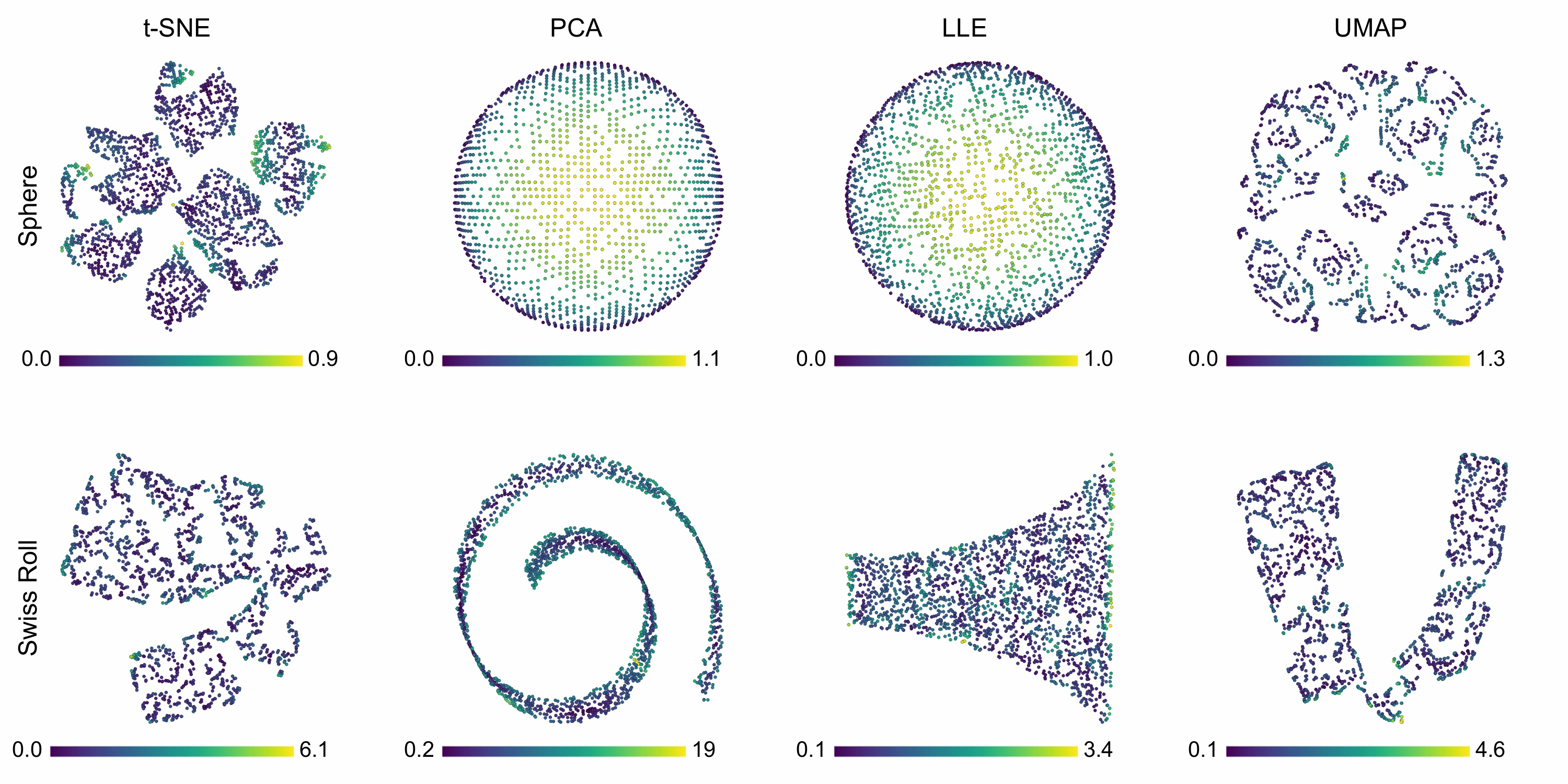}
\caption
{
Validation maps showing inverse-projection error. Top row: Validation of the sphere dataset under four different projections.
Bottom row: Validation of the swiss roll dataset under the same projections.
Reconstruction error is computed by comparing the inverse-projected location with the original point's location.
Each figure is scaled separately to highlight how the error distributed within that figure, so figures cannot be directly compared.
The min and max error of each figure can be found below each plot on either end of a color legend.
}
\label{fig:validation_table}
\end{figure*}

We explore \sys's performance on two well-understood synthetic datasets, Sphere and Swiss Roll (Sec.~\ref{sec:method:data}).
Simple datasets where the projections are well understood give us greater ability to reason about the inverse projection.
In particular it is easier to understand how error is distributed across a dataset, as well as which projections will incur higher error.
To illustrate this, we once again split the datasets into training ($D_s$) and test sets ($D_p$), this time having $75$ and $25$ percent of the total data, respectively.
We then plot the projections of test portion ($P_p$) of these two datasets in Fig.~\ref{fig:validation_table} and color the points by the root-mean-squared error between the inverse-projections ($B(P_p)$) and the true high-dimensional data ($D_p$).
Error colormaps are normalized within each image, so that we can better see error variations \emph{within} a given projection.
Hence, colors cannot be compared across rows  or columns of Fig.~\ref{fig:validation_table}.

When analyzing inverse projection results, we must remember that the concept of error encompasses inaccuracies and faults in \emph{both} the projection and the  inverse-projection methods. 
For example, linear techniques like PCA will have a substantially different error profile compared to non-linear techniques such as t-SNE, LLE, or UMAP.
For the sphere, t-SNE and UMAP are able to peel away the surface, and error seems to congregate along the edges of the structures that make up the peel.
In contrast, PCA and LLE end up with a slice out of the sphere causing the largest  error in the center of their slices.
For the Swiss Roll dataset, t-SNE, LLE, and UMAP are able to remove the swirl, with UMAP and t-SNE making similar ribboned shapes and LLE unraveling to an rectangle with perspective. In contrast, PCA keeps the general shape of the spiral, causing a speckling of high error throughout the whole structure.
Projecting high-dimensional space down to 2D is inherently lossy, and each method will project the data to 2D differently.
This difference is not only visual -- each projection method emphasizes certain aspects of the data. 
As a result, different techniques throw away different portions and amounts of the high-dimensional data when performing the projection.
This means that certain projection techniques will be easier to inverse-project than others. 
For example, PCA does not aim to prevent overdrawing or projecting different points to the same two-dimensional location. As such, several data points can be projected to the same position in 2D space, making it impossible to correctly learn an inverse.
In contrast, t-SNE and other non-linear techniques work to maintain local neighborhood relationships; when projecting a set of points, they try to preserve the relative distances in the projection that exist in the original space. In cases where there is poor preservation of inter-cluster distances, \sys remains a valuable tool.
If an area in the projection is shrunk or expanded relative to the high-dimensional space, the rate of change between inverse-projections will either increase or decrease respectively.
When the interpolation moves very quickly,  \sys may be less useful for tasks like dynamic imputation (Sec.~\ref{sec:application:reconstruction}), but \sys can help identify these spots with gradient maps (Sec.~\ref{sec:application:model_gradient}).
The properties of each projection technique inform and define the types of errors exhibited during the inverse-projection process.
As Fig.~\ref{fig:validation_table} shows, one consequence of PCA projecting multiple distant points to a small region on a 2D  plane is that the inverse-projected points  will likely be erroneous.
In this case, the error increases as the distance in the high-dimensional space increases between points co-located on the 2D  projection. 
Conversely, for t-SNE and UMAP, the non-linear projections distort the input geometry, often into shapes that no longer resemble the topology of the original data. 
In return, the inverse-projected data points from 2D back to the high-dimensional space are much closer to their original positions, resulting in significantly smaller total error.
In other words, better grouping by similarity as well as better separation of points will make inverse-projection easier.

\subsection{Dense map of inverse projection error}
\label{sec:evaluation:invmap}
\begin{figure*}[h]
\centering
\includegraphics[width=0.8\textwidth]{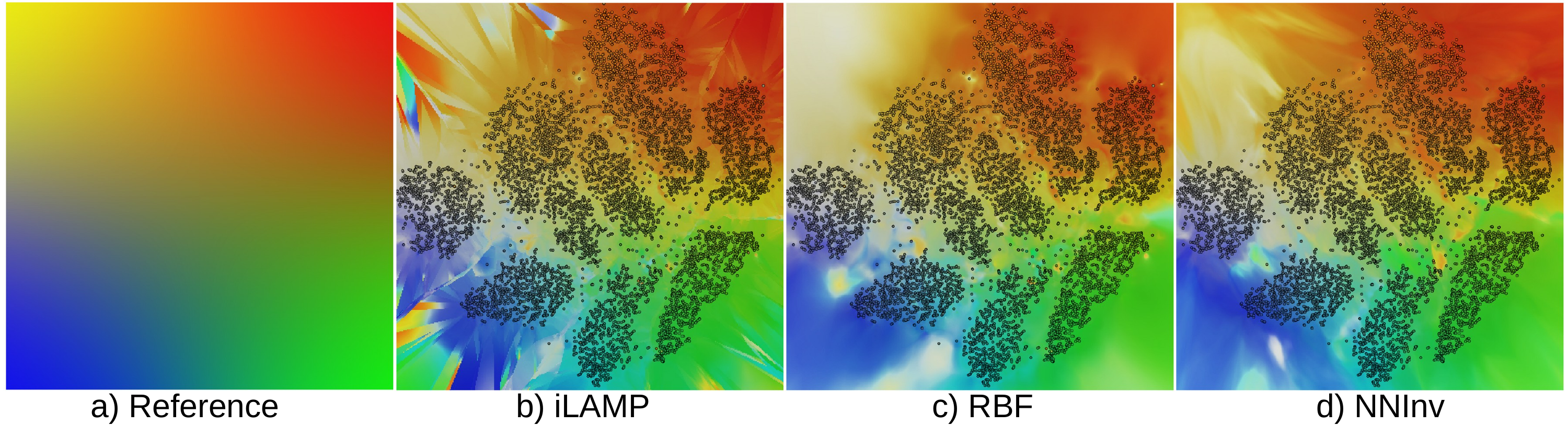}
\caption{Differences of ``round-trip errors'' with three inverse techniques (iLAMP, RBF, and NNinv). See Sec.~\ref{sec:evaluation:invmap} for an in-depth discussion.
}
\label{fig:invmap}
\end{figure*}
%
%

Evaluation of inverse-projection methods often uses error metrics defined for direct projections such as stress or reprojection error\cite{amorim:2012:ilamp, amorim:2015:rbf}.
However, the above metrics only gauge the error at the locations of projection points $P(D)$. The same is actually the case for all errors for direct projections we are aware of -- they only gauge how good a (direct) projection is at the locations of the scatterplot points. As explained earlier in Sec.~\ref{sec:related_work:back_projection}, the key use-case of inverse projections is the \emph{out-of-sample} one, where one inversely projects \emph{different} points than $P(D)$. 
We next propose a validation approach that considers the out-of-sample case, \emph{i.e.}, evaluates the quality of $B$ at all points in $\mathbb{R}^2$. We proceed as follows. Given a dataset $D$, we construct $P(D)$ as usual given a user-chosen projection technique $P$, and use $(D,P(D))$ to train our inverse projection $B$. Next, we discretize the projection space $\mathbb{R}^2$ using a pixel grid with a given resolution $R$, in our case $R = 400$. Then, for every pixel $\mathbf{y}$, we compute the pixel $\mathbf{y}' =  P(B(\mathbf{y}))$ given by the ``round trip'' of back projecting it to $\mathbb{R}^d$ and next projecting it again to $\mathbb{R}^2$. To perform this, we must assume that $P$ is \emph{parametric}. Then, ideally, $\mathbf{y}=\mathbf{y}'$ for all pixels $\mathbf{y}$. This way, we can assess an inverse projection error also for points in $\mathbb{R}^2$ which do not correspond to projections of points in our given dataset $D$.
We visualize the round-trip errors as a dense map as follows. We create a hue image by bilinear interpolation of four different hues (Fig.~\ref{fig:invmap}a). Next, we color every pixel $\mathbf{y}$ by the hue of the round-trip pixel $\mathbf{y}'$ and set its luminance to $\|\mathbf{y} - \mathbf{y}'\|$. Dense map areas which show the same color gradient as Fig.~\ref{fig:invmap}a have, thus, low inverse-projection errors. Bright areas and/or hue differences from this gradient show large projection errors. Scatterplot points are colored in the same way, but use a slightly lower brightness value to avoid confusion with the map pixels. Figures~\ref{fig:invmap}b-d show the error maps for iLAMP, RBF, and \sys for the inverse projection of the MNIST dataset projected by t-SNE. We see that \sys creates a color gradient which is close to the reference one, has minimal discontinuities, and has few bright spots. Hence, \sys can inverse-project the entire 2D space without introducing large amounts of error.

\subsection{Scalability in Training and Inference}
\label{sec:evaluation:scalability}
Scalability implies the effort required to \emph{train} our method and, separately, the effort needed to \emph{infer} $B(T)$ as function of the size of the dataset $Y$ to inversely project. 
Concerning training, Table~\ref{tab:epochs} shows the number of training epochs needed to obtain convergence (defined as in Sec.~\ref{sec:method:practicalities}) as function of the training set size $|D_s|$, for all three considered datasets and $P=\mbox{t-SNE}$.
Columns 2..4 indicate averages for multiple runs created by randomly sampling $D_s$ from the entire dataset $D$.
Overall, we obtain convergence for roughly 150 epochs for all datasets and training-set sizes. 


\begin{table}[htb]
\centering
\begin{tabular}{| c | c | c | c | c |}
\hline
Training set  &   \multicolumn{3}{c |}{Average \# epochs for each dataset $D$}            & Row  \\ \cline{2-4}
    size $|D_s|$             & Blobs & Fashion-MNIST & MNIST   & averages  \\ \hline
500           & 268.0        & 214.0                & 213.5        & 192.5 \\
1000          & 190.5        & 129.0                & 147.5        & 149.0 \\
2000          & 153.0        & 112.0                & 111.0        & 112.5 \\
5000          & 103.0        & 120.5                & 138.0        & 127.5 \\
7000          & 127.0        & 118.5                & 151.0        & 144.0 \\
10000         & 82.0         & 124.5                & 142.5        & 146.5 \\ \hline
column avg   & 153.9        & 136.4                & 150.6        & 145.3 \\ \hline
\end{tabular}
\scriptsize
\caption{Training effort until convergence.}
\label{tab:epochs}
\end{table}

\begin{figure}[!htp!]
\centering
\includegraphics[width=1.0\linewidth]{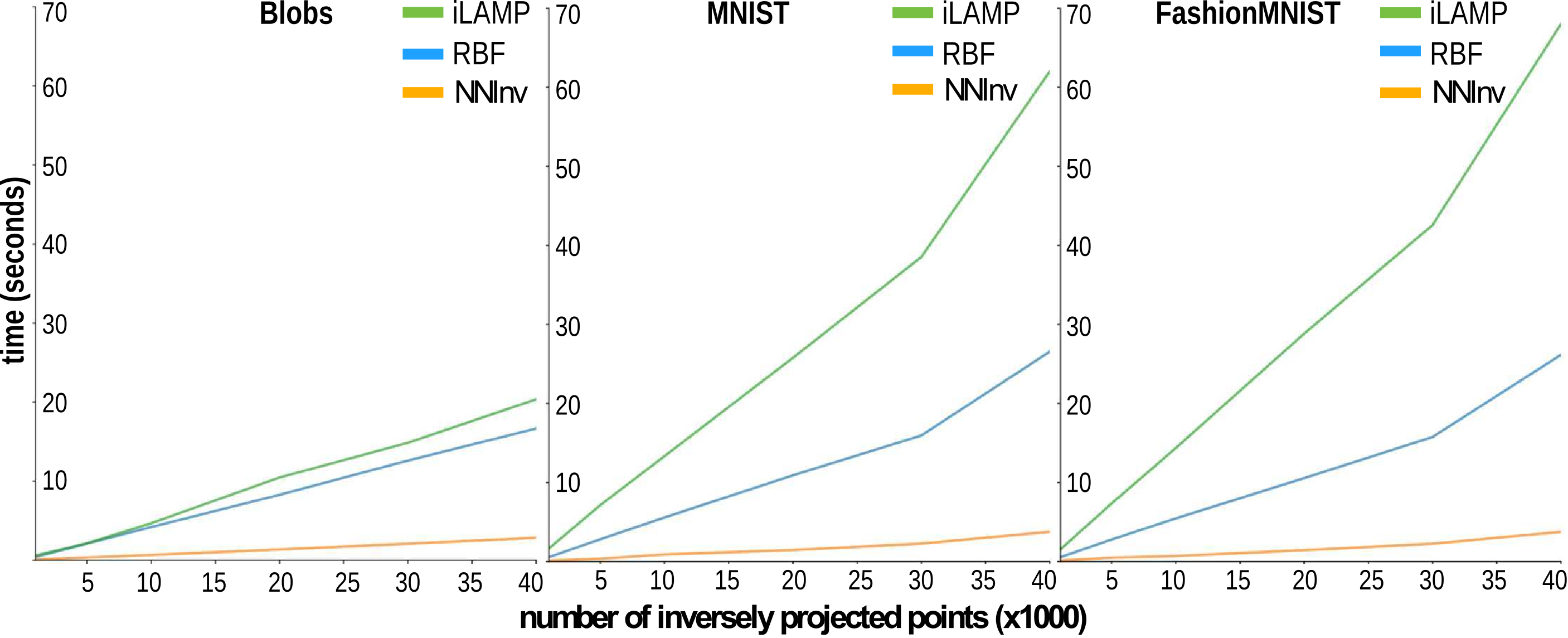}
\vspace{-0.2cm}
\caption{Inverse-projection speed \emph{vs}  number of samples\,\cite{espadoto:2019:nn_inv}. Inverse-projection speed is of great importance for applications that require the inverse-projection of many samples, as is the case for the aforementioned classifier and gradient maps (Secs.~\ref{sec:application:model_agreements} and~\ref{sec:application:model_gradient}).}
\vspace{-0.2cm}
\label{fig:time_eval}
\end{figure}

Figure~\ref{fig:time_eval} shows the inference speed for all three datasets. Speed  does not depend on the projection method $P$ -- once \sys is trained, its performance is linear in the number of inversely-projected samples. When computing inference speed, we inversely project \emph{any} point in $\mathbb{R}^2$ and not just points in $P(D)$. Indeed, for assessing speed, we do not need ground-truth information. Moreover, in real use cases, one would inversely project \emph{unseen} data, for which such ground-truth information is not available.
We see that both RBF and iLAMP have a superlinear behavior, while \sys (our method) is basically linear. \sys is roughly one magnitude order faster than RBF and nearly two magnitude orders faster than iLAMP for 40K samples or more. This speed-up is crucial for applications that need to inversely project hundreds of thousands of samples (or more), like in the construction of dense maps\,\cite{ rodrigues:2018:classifier_boundaries, espadoto:2019:nn_inv} and the maps in Sec.~\ref{sec:application:model_agreements} and~\ref{sec:application:model_gradient}. 
\sys constructs such maps in \emph{seconds}, while iLAMP and RBF need (tens of) minutes, making human-in-the-loop usage of such methods impossible in visual analytics scenarios -- one of the key reasons why dense maps are built in the first place. This scalability is one of the most important advantages of \sys.

All experiments were run on a 4-core Intel E3-1240 v6 at 3.7 GHz with 64 GB RAM and an NVidia GeForce GTX 1070 GPU with 8 GB VRAM, and the code was implemented in Python 3 using the Keras library (\url{keras.io}).



\section{Limitations}
\label{sec:limitations}
\sys is scalable, accurate, and relatively smooth, as shown in Sec.~\ref{sec:evaluation}.
Yet, using a neural network does have its disadvantages\cite{bengio:2013:deep}.
A neural network (1) requires a particular threshold of good quality training examples, (2) can be computationally expensive to train, and (3) can be generally hard to interpret. 
In Sec.~\ref{sec:evaluation:quality} we show acceptable mean squared error with as few as 500 training examples, and caution that below that threshold, our technique will not perform as successfully.
In all of the examples in this paper, the projections \sys are trained on are \emph{good quality} projections, obtained by choosing reasonable values for the projection's hyperparameters.
Good quality projections are generally more likely to have the qualities (as described in Sec.~\ref{sec:evaluation:qualitative}) required for accurate inverse-projection.
While \sys is useful in helping to interpret projections (\emph{e.g.}, Fig.~\ref{fig:interpolation_table}), it can be difficult to reason about \sys itself, since neural networks are hard to interpret in general.
That is, our metrics show that \sys performs better as it can approximate nonlinear patterns, but it is not obvious how \sys does this.
We leave the explainability of \sys's improved performance to future work.

\section{Discussion and Future Work}
\label{sec:discussion}
Future inverse-projection research can take several interesting directions.
Of particular relevance is the discussion in Sec.~\ref{sec:evaluation:qualitative} regarding the properties of projection techniques and their inverses.
As Sec.~\ref{sec:evaluation:qualitative} shows, when discussing the invertibility of projection functions, we find that not all projection methods are equally suitable for the inverse-projection method: PCA is worse than t-SNE or UMAP because multiple data points can be projected into the same 2D pixel.
Interestingly, a type of projection that is designed specifically for its invertibility is the encoder portion of an autoencoder.
When trained together with the decoder, the entire process optimizes for the recoverability of data points from input to output.
Yet, a user would have a hard time understanding the embedding of a regular encoder because there is no intentionally designed structure in the embedding space created with an encoder. Also, there are no guarantees about neighborhood preservation or relative distance preservation.
The tradeoff between understandability of the latent space created by a projection and the appropriateness of the projection for learning its inverse is interesting.
On one hand, a projection technique may sacrifice some information to create a more insightful, or more spatially intuitive,  visualization. 
Yet, the use of inverse-projection can lead to novel visualization and interaction techniques that can better help the user explore and understand a high-dimensional space.
Further steps should be taken to find a happy medium between these two extremes, whether that be autoencoders with some cost for occlusion, or spacing items too far apart, or a projection technique with a greater loss for discarding information.
%

%
Accessible and fast inverse projections will have far-reaching impacts on visual analytics (VA) systems that use projections.
We believe that a deep learning approach to inverse projection is especially accessible given today's robust ecosystem for neural network development\,\cite{chollet:2015:keras, google:2019:kerastuner, abadi:2015:tensorflow, paszke:2019:pytorch}.
We hope that future works along this line of research continue to leverage approachable methods and libraries that ease adoption for tool builders.
The most potential use-case is hypothesis generation made possible by dynamic imputation (Sec.~\ref{sec:application:reconstruction}), but several different augmentations exist, \emph{e.g.}, adding extra information showing how models understand the data space in the same vein as classifier agreement maps (Sec.~\ref{sec:application:model_agreements}), or helping projection users to better understand the underlying structure as in gradient maps (Sec.~\ref{sec:application:model_gradient}).
We are particularly interested in how combinations of these techniques, as the hypothesis generation paired with gradient map style backgrounds, can help users who are less familiar with projection techniques make sense of overview projections in VA applications.
%

%
Lastly, we believe there are several applications of this technique that should be explored further. Projections and inverse-projections can be used to explore the space of different 2D charts that have themselves been projected to 2D (in a manner similar to ChartSeer\cite{zhao:2020:chartseer}), and data that is often modeled on graphs, such as molecular data.
\section{Conclusion}
\label{sec:conclusions}
%
%
%
%
%
%

%
In this paper, we present \sys, a deep learning approach to learning the inverse of projection functions.
Similar to existing works such as iLAMP and RBF, \sys is agnostic of the projection used, \emph{i.e.}, it can learn to invert any projection algorithm (such as PCA, t-SNE, UMAP, LLE, etc.).
\sys uses a trained neural network to learn the approximate mapping from a given 2D scatterplot produced by a projection algorithm to the corresponding high-dimensional data.
We find that \sys can be more accurate than iLAMP and RBF on both synthetic and real-world datasets, and is more scalable to large datasets: Once trained, \sys can perform inferencing within less than 10 milliseconds  when running in a browser on a laptop, which makes \sys a more suitable technique than iLAMP and RBF for interactive visualizations.
%
%
%
%
%
%
Lastly, we show the potential of \sys for analysis tasks such as hypothesis generation, classifier agreement, and gradient visualization. These are three areas important to the field of visual analytics and serve as evidence to the possibility of the broad applicability of \sys in high-dimensional data exploration and analysis.
\ifCLASSOPTIONcompsoc
  \section*{Acknowledgments}
\else
  \section*{Acknowledgment}
\fi

This work was supported by National Science Foundation grants IIS1452977, OAC-1940175, OAC-1939945, DGE-1855886, DARPA grant FA8750-17-2-0107, and DOD grant HQ0860-20-C-7137. We would also like to thank the reviewers for their helpful feedback.

\ifCLASSOPTIONcaptionsoff
  \newpage
\fi



%

\bibliographystyle{IEEEtran}
\bibliography{unprojection}

\begin{thebibliography}{10}
\providecommand{\url}[1]{#1}
\csname url@samestyle\endcsname
\providecommand{\newblock}{\relax}
\providecommand{\bibinfo}[2]{#2}
\providecommand{\BIBentrySTDinterwordspacing}{\spaceskip=0pt\relax}
\providecommand{\BIBentryALTinterwordstretchfactor}{4}
\providecommand{\BIBentryALTinterwordspacing}{\spaceskip=\fontdimen2\font plus
\BIBentryALTinterwordstretchfactor\fontdimen3\font minus
  \fontdimen4\font\relax}
\providecommand{\BIBforeignlanguage}[2]{{%
\expandafter\ifx\csname l@#1\endcsname\relax
\typeout{** WARNING: IEEEtran.bst: No hyphenation pattern has been}%
\typeout{** loaded for the language `#1'. Using the pattern for}%
\typeout{** the default language instead.}%
\else
\language=\csname l@#1\endcsname
\fi
#2}}
\providecommand{\BIBdecl}{\relax}
\BIBdecl

\bibitem{buhlmann:2011:stats_for_high_dim}
P.~Buhlmann, \emph{\BIBforeignlanguage{eng}{Statistics for high-dimensional
  data : methods, theory and applications}}, ser. Springer series in
  statistics.\hskip 1em plus 0.5em minus 0.4em\relax Heidelberg ; New York:
  Springer, 2011.

\bibitem{maaten:2008:tsne}
L.~van~der Maaten and G.~Hinton, ``Visualizing data using t-sne,''
  \emph{Journal of Machine Learning Research}, vol.~9, no.~86, pp. 2579--2605,
  2008.

\bibitem{pearson:1901:pca}
K.~Pearson, ``{LIII}. on lines and planes of closest fit to systems of points
  in space,'' \emph{The London, Edinburgh, and Dublin Philosophical Magazine
  and Journal of Science}, vol.~2, no.~11, pp. 559--572, 1901.

\bibitem{roweis:2000:lle}
S.~T. Roweis and L.~K. Saul, ``Nonlinear dimensionality reduction by locally
  linear embedding,'' \emph{science}, vol. 290, no. 5500, pp. 2323--2326, 2000.

\bibitem{mcinnes:2018:umap}
L.~{McInnes}, J.~{Healy}, and J.~{Melville}, ``{UMAP: Uniform Manifold
  Approximation and Projection for Dimension Reduction},'' \emph{ArXiv
  e-prints}, Feb. 2018.

\bibitem{espadoto19}
M.~Espadoto, R.~M. Martins, A.~Kerren, N.~S.~T. Hirata, and A.~C. Telea,
  ``Toward a quantitative survey of dimension reduction techniques,''
  \emph{IEEE Transactions on Visualization and Computer Graphics}, vol.~27,
  no.~3, pp. 2153--2173, 2021.

\bibitem{Li:2017:application_of_tsne}
W.~Li, J.~E. Cerise, Y.~Yang, and H.~Han, ``Application of t-sne to human
  genetic data,'' \emph{Journal of bioinformatics and computational biology},
  vol. 15 4, p. 1750017, 2017.

\bibitem{martins14}
R.~Martins, D.~Coimbra, R.~Minghim, and A.~C. Telea, ``Visual analysis of
  dimensionality reduction quality for parameterized projections,''
  \emph{Computers \& Graphics}, vol.~41, pp. 26--42, 2014.

\bibitem{nonato18}
L.~Nonato and M.~Aupetit, ``Multidimensional projection for visual analytics:
  Linking techniques with distortions, tasks, and layout enrichment,''
  \emph{IEEE TVCG}, 2018.

\bibitem{amorim:2012:ilamp}
E.~P. {dos Santos Amorim}, E.~V. {Brazil}, J.~{Daniels}, P.~{Joia}, L.~G.
  {Nonato}, and M.~C. {Sousa}, ``{iLAMP}: Exploring high-dimensional spacing
  through backward multidimensional projection,'' in \emph{2012 IEEE Conference
  on Visual Analytics Science and Technology (VAST)}, 2012, pp. 53--62.

\bibitem{rodrigues:2019:classifier_boundaries}
F.~C.~M. Rodrigues, M.~Espadoto, R.~Hirata, and A.~C. Telea, ``Constructing and
  visualizing high-quality classifier decision boundary maps,''
  \emph{Information}, vol.~10, no.~9, p. 280, Sep 2019.

\bibitem{amorim:2015:rbf}
E.~Amorim, E.~V. Brazil], J.~Mena-Chalco, L.~Velho, L.~G. Nonato, F.~Samavati,
  and M.~C. Sousa], ``Facing the high-dimensions: Inverse projection with
  radial basis functions,'' \emph{Computers \& Graphics}, vol.~48, pp. 35 --
  47, 2015.

\bibitem{espadoto:2020:innp}
M.~Espadoto, N.~S.~T. Hirata, A.~X. Falc{\~{a}}o, and A.~C. Telea, ``Improving
  neural network-based multidimensional projections,'' in \emph{Proceedings of
  the 15th International Joint Conference on Computer Vision, Imaging and
  Computer Graphics Theory and Applications}, vol.~3.\hskip 1em plus 0.5em
  minus 0.4em\relax {SCITEPRESS}, 2020, pp. 29--41.

\bibitem{scikitboundary}
Scikit\-Learn.org, ``Classifier comparison,''
  \url{https://scikit-learn.org/stable/auto_examples/classification/plot_classifier_comparison.html},
  retrieved April 30, 2020.

\bibitem{andrews:1972:plot_greater_than_2d}
D.~F. Andrews, ``Plots of high-dimensional data,'' \emph{Biometrics}, vol.~28,
  no.~1, pp. 125--136, 1972.

\bibitem{geng:2013:3d_display}
J.~Geng, ``Three-dimensional display technologies,'' \emph{Advances in optics
  and photonics}, vol.~5, no.~4, pp. 456--535, 2013.

\bibitem{gorban:2008:principal_manifold_techniques}
A.~N. Gorban, B.~K{\'e}gl, D.~C. Wunsch, A.~Y. Zinovyev \emph{et~al.},
  \emph{Principal manifolds for data visualization and dimension
  reduction}.\hskip 1em plus 0.5em minus 0.4em\relax Springer, 2008, vol.~58.

\bibitem{van:2009:dim_reduction_survey}
L.~Van Der~Maaten, E.~Postma, and J.~Van~den Herik, ``Dimensionality reduction:
  a comparative review,'' \emph{J Mach Learn Res}, vol.~10, no. 66-71, p.~13,
  2009.

\bibitem{joia:2011:lamp}
P.~{Joia}, D.~{Coimbra}, J.~A. {Cuminato}, F.~V. {Paulovich}, and L.~G.
  {Nonato}, ``Local affine multidimensional projection,'' \emph{IEEE
  Transactions on Visualization and Computer Graphics}, vol.~17, no.~12, pp.
  2563--2571, 2011.

\bibitem{silva:2012:user_centered_projection}
C.~T. {Silva}, F.~V. {Paulovich}, and L.~G. {Nonato}, ``User-centered
  multidimensional projection techniques,'' \emph{Computing in Science
  Engineering}, vol.~14, no.~4, pp. 74--81, 2012.

\bibitem{sorzano:2014:dim_reduction_survey}
C.~O.~S. Sorzano, J.~Vargas, and A.~P. Montano, ``A survey of dimensionality
  reduction techniques,'' \emph{arXiv preprint arXiv:1403.2877}, 2014.

\bibitem{sacha:2016:dim_reduction_interaction_survey}
D.~Sacha, L.~Zhang, M.~Sedlmair, J.~A. Lee, J.~Peltonen, D.~Weiskopf, S.~C.
  North, and D.~A. Keim, ``Visual interaction with dimensionality reduction: A
  structured literature analysis,'' \emph{IEEE transactions on visualization
  and computer graphics}, vol.~23, no.~1, pp. 241--250, 2016.

\bibitem{jeong:2009:ipca}
D.~H. Jeong, C.~Ziemkiewicz, B.~Fisher, W.~Ribarsky, and R.~Chang, ``{iPCA}: An
  interactive system for pca-based visual analytics,'' in \emph{Computer
  Graphics Forum}, vol.~28, no.~3.\hskip 1em plus 0.5em minus 0.4em\relax Wiley
  Online Library, 2009, pp. 767--774.

\bibitem{paulovich:2008:least}
F.~V. Paulovich, L.~G. Nonato, R.~Minghim, and H.~Levkowitz, ``Least square
  projection: A fast high-precision multidimensional projection technique and
  its application to document mapping,'' \emph{IEEE Transactions on
  Visualization and Computer Graphics}, vol.~14, no.~3, pp. 564--575, 2008.

\bibitem{paulovich2006text}
F.~V. Paulovich and R.~Minghim, ``Text map explorer: a tool to create and
  explore document maps,'' in \emph{Proc. IEEE IV}, 2006, pp. 245--251.

\bibitem{cunningham:2015:linear_dim_reduction_survey}
J.~P. Cunningham and Z.~Ghahramani, ``Linear dimensionality reduction: Survey,
  insights, and generalizations,'' \emph{Journal of Machine Learning Research},
  vol.~16, no.~89, pp. 2859--2900, 2015.

\bibitem{yin07_survey}
H.~Yin, ``Nonlinear dimensionality reduction and data visualization: A
  review,'' \emph{Intl. Journal of Automation and Computing}, vol.~4, no.~3,
  pp. 294--303, 2007.

\bibitem{hoffman:1997:radviz}
P.~{Hoffman}, G.~{Grinstein}, K.~{Marx}, I.~{Grosse}, and E.~{Stanley}, ``Dna
  visual and analytic data mining,'' in \emph{Proceedings. Visualization '97
  (Cat. No. 97CB36155)}, 1997, pp. 437--441.

\bibitem{angelini:2019:enhancing_radviz}
M.~Angelini, G.~Blasilli, S.~Lenti, A.~Palleschi, and G.~Santucci, ``Towards
  enhancing radviz analysis and interpretation,'' in \emph{2019 IEEE
  Visualization Conference (VIS)}.\hskip 1em plus 0.5em minus 0.4em\relax IEEE,
  2019, pp. 226--230.

\bibitem{pagliosa:2019:radviz++}
L.~d.~C. Pagliosa and A.~C. Telea, ``Radviz++: Improvements on radial-based
  visualizations,'' in \emph{Informatics}, vol.~6.\hskip 1em plus 0.5em minus
  0.4em\relax Multidisciplinary Digital Publishing Institute, 2019, p.~16.

\bibitem{hoffman02}
P.~Hoffman and G.~Grinstein, ``A survey of visualizations for high-dimensional
  data mining,'' \emph{Information Visualization in Data Mining and Knowledge
  Discovery}, vol. 104, pp. 47--82, 2002.

\bibitem{bunte11}
K.~Bunte, M.~Biehl, and B.~Hammer, ``A general framework for dimensionality
  reducing data visualization mapping,'' \emph{Neural Computation}, vol.~24,
  no.~3, pp. 771--804, 2012.

\bibitem{maljovec15}
S.~Liu, D.~Maljovec, B.~Wang, P.-T. Bremer, and V.~Pascucci, ``Visualizing
  high-dimensional data: Advances in the past decade,'' \emph{IEEE TVCG},
  vol.~23, no.~3, pp. 1249--1268, 2015.

\bibitem{seifert:2010:stress_maps}
C.~Seifert, V.~Sabol, and W.~Kienreich, ``Stress maps: Analysing local
  phenomena in dimensionality reduction based visualisations.'' in
  \emph{EuroVAST@ EuroVis}, 2010.

\bibitem{venna10}
J.~Venna, J.~Peltonen, K.~Nybo, H.~Aidos, and S.~Kaski, ``Information retrieval
  perspective to nonlinear dimensionality reduction for data visualization,''
  \emph{JMLR}, vol.~11, pp. 451--490, 2010.

\bibitem{martins15}
R.~Martins, R.~Minghim, and A.~C. Telea, ``Explaining neighborhood preservation
  for multidimensional projections,'' in \emph{Proc. CGVC}.\hskip 1em plus
  0.5em minus 0.4em\relax Eurographics, 2015, pp. 121--128.

\bibitem{aupetit:2007:visualizing_distortions_in_projections}
M.~Aupetit, ``Visualizing distortions and recovering topology in continuous
  projection techniques,'' \emph{Neurocomputing}, vol.~70, no. 7-9, pp.
  1304--1330, 2007.

\bibitem{lespinats:2011:checkviz}
S.~Lespinats and M.~Aupetit, ``Checkviz: Sanity check and topological clues for
  linear and non-linear mappings,'' \emph{Computer Graphics Forum}, vol.~30,
  no.~1, pp. 113--125, 2011.

\bibitem{faust:2018:dim_reader}
R.~Faust, D.~Glickenstein, and C.~Scheidegger, ``Dimreader: Axis lines that
  explain non-linear projections,'' \emph{IEEE transactions on visualization
  and computer graphics}, vol.~25, no.~1, pp. 481--490, 2018.

\bibitem{kerren:2020:tvisne}
A.~Chatzimparmpas, R.~M. Martins, and A.~Kerren, ``t-visne: Interactive
  assessment and interpretation of t-sne projections,'' \emph{IEEE Transactions
  on Visualization \& Computer Graphics}, vol.~26, no.~08, pp. 2696--2714, aug
  2020.

\bibitem{stahnke:2015:probing_projections}
J.~Stahnke, M.~D{\"o}rk, B.~M{\"u}ller, and A.~Thom, ``Probing projections:
  Interaction techniques for interpreting arrangements and errors of
  dimensionality reductions,'' \emph{IEEE transactions on visualization and
  computer graphics}, vol.~22, no.~1, pp. 629--638, 2015.

\bibitem{brown:2012:dis}
E.~T. Brown, J.~Liu, C.~E. Brodley, and R.~Chang, ``Dis-function: Learning
  distance functions interactively,'' in \emph{Visual Analytics Science and
  Technology}.\hskip 1em plus 0.5em minus 0.4em\relax IEEE, 2012, pp. 83--92.

\bibitem{dowling:2019:sirius}
M.~{Dowling}, J.~{Wenskovitch}, J.~T. {Fry}, S.~{Leman}, L.~{House}, and
  C.~{North}, ``Sirius: Dual, symmetric, interactive dimension reductions,''
  \emph{IEEE Transactions on Visualization and Computer Graphics}, vol.~25,
  no.~1, pp. 172--182, 2019.

\bibitem{nnp}
M.~Espadoto, N.~S.~T. Hirata, and A.~C. Telea, ``Deep learning multidimensional
  projections,'' \emph{Information Visualization}, vol.~19, no.~3, pp.
  247--269, 2020.

\bibitem{rodrigues:2018:classifier_boundaries}
F.~C. {M. Rodrigues}, R.~{Hirata}, and A.~C. {Telea}, ``Image-based
  visualization of classifier decision boundaries,'' in \emph{SIBGRAPI
  Conference on Graphics, Patterns and Images}, 2018, pp. 353--360.

\bibitem{hinton2006reducing}
G.~E. Hinton and R.~R. Salakhutdinov, ``Reducing the dimensionality of data
  with neural networks,'' \emph{Science}, vol. 313, no. 5786, pp. 504--507,
  2006.

\bibitem{vernier20}
E.~Vernier, R.~Garcia, I.~da~Silva, J.~Comba, and A.~Telea, ``Quantitative
  evaluation of time-dependent multidimensional projection techniques,''
  \emph{Computer Graphics Forum}, vol.~39, no.~3, 2020.

\bibitem{mamani:2013:user_driven_feature_space_transformation}
G.~M.~H. Mamani, F.~M. Fatore, L.~G. Nonato, and F.~V. Paulovich, ``User-driven
  feature space transformation,'' \emph{Computer Graphics Forum}, vol.~32, no.
  3pt3, pp. 291--299, 2013.

\bibitem{kriegeskorte:2012:inverse_mds}
N.~Kriegeskorte and M.~Mur, ``Inverse mds: Inferring dissimilarity structure
  from multiple item arrangements,'' \emph{Frontiers in Psychology}, vol.~3, p.
  245, 2012.

\bibitem{cavallo:2018:praxis}
M.~Cavallo and c.~Demiralp, ``A visual interaction framework for dimensionality
  reduction based data exploration,'' in \emph{Conference on Human Factors in
  Computing Systems}, ser. CHI ’18.\hskip 1em plus 0.5em minus 0.4em\relax
  New York, NY, USA: Association for Computing Machinery, 2018, p. 1–13.

\bibitem{zhao:2020:chartseer}
J.~Zhao, M.~Fan, and M.~Feng, ``Chartseer: Interactive steering exploratory
  visual analysis with machine intelligence,'' \emph{IEEE Transactions on
  Visualization and Computer Graphics}, pp. 1--1, 2020.

\bibitem{kusner2017grammar}
M.~J. Kusner, B.~Paige, and J.~M. Hern{\'a}ndez-Lobato, ``Grammar variational
  autoencoder,'' in \emph{International Conference on Machine Learning}.\hskip
  1em plus 0.5em minus 0.4em\relax PMLR, 2017, pp. 1945--1954.

\bibitem{goodfellow19}
I.~Goodfellow, Y.~Bengio, and A.~Courville, \emph{Deep Learning}.\hskip 1em
  plus 0.5em minus 0.4em\relax MIT Press, 2017.

\bibitem{liu:2019:latent_space_catography}
Y.~Liu, E.~Jun, Q.~Li, and J.~Heer, ``Latent space cartography: Visual analysis
  of vector space embeddings,'' in \emph{Computer Graphics Forum},
  vol.~38.\hskip 1em plus 0.5em minus 0.4em\relax Wiley Online Library, 2019,
  pp. 67--78.

\bibitem{spinner:2018:towards_interpretable_latent_space}
T.~Spinner, J.~K{\"o}rner, J.~G{\"o}rtler, and O.~Deussen, ``Towards an
  interpretable latent space: an intuitive comparison of autoencoders with
  variational autoencoders,'' in \emph{Proceedings of the Workshop on
  Visualization for AI Explainability 2018}, 2018.

\bibitem{higgins:2017:beta_vae}
I.~Higgins, L.~Matthey, A.~Pal, C.~Burgess, X.~Glorot, M.~Botvinick,
  S.~Mohamed, and A.~Lerchner, ``beta-vae: Learning basic visual concepts with
  a constrained variational framework,'' in \emph{ICLR}, 2017.

\bibitem{kim:2019:disentangling_factorising}
H.~Kim and A.~Mnih, ``Disentangling by factorising,'' 2019.

\bibitem{chen:2019:isolating_disentanglement_vae}
T.~Q. Chen, X.~Li, R.~B. Grosse, and D.~Duvenaud, ``Isolating sources of
  disentanglement in variational autoencoders,'' \emph{CoRR}, vol.
  abs/1802.04942, 2018.

\bibitem{Gou:2020:valtd}
L.~{Gou}, L.~{Zou}, N.~{Li}, M.~{Hofmann}, A.~K. {Shekar}, A.~{Wendt}, and
  L.~{Ren}, ``Vatld: A visual analytics system to assess, understand and
  improve traffic light detection,'' \emph{IEEE Transactions on Visualization
  and Computer Graphics}, pp. 1--1, 2020.

\bibitem{garcia18}
R.~Garcia, A.~Telea, B.~da~Silva, J.~Torresen, and J.~Comba, ``A
  task-and-technique centered survey on visual analytics for deep learning
  model engineering,'' \emph{Computers and Graphics}, vol.~77, pp. 30--49,
  2018.

\bibitem{lecun:2010:mnist}
Y.~LeCun, C.~Cortes, and C.~Burges, ``Mnist handwritten digit database,''
  \emph{ATT Labs [Online]. Available: http://yann. lecun. com/exdb/mnist},
  vol.~2, 2010.

\bibitem{espadoto:2019:nn_inv}
M.~Espadoto, F.~C.~M. Rodrigues, N.~S.~T. Hirata, R.~Hirata~Jr., and A.~C.
  Telea, ``{Deep Learning Inverse Multidimensional Projections},'' in
  \emph{EuroVis Workshop on Visual Analytics}, T.~v. Landesberger and
  C.~Turkay, Eds.\hskip 1em plus 0.5em minus 0.4em\relax The Eurographics
  Association, 2019.

\bibitem{xiao:2017:fashion_mnist}
H.~Xiao, K.~Rasul, and R.~Vollgraf, ``Fashion-mnist: a novel image dataset for
  benchmarking machine learning algorithms,'' \emph{CoRR}, vol. abs/1708.07747,
  2017.

\bibitem{balasubramanian2002isomap}
M.~Balasubramanian and E.~L. Schwartz, ``The {Isomap} algorithm and topological
  stability,'' \emph{Science}, vol. 295, no. 5552, pp. 7--7, 2002.

\bibitem{elsken:2019:nas_survey}
T.~Elsken, J.~H. Metzen, and F.~Hutter, ``Neural architecture search: A
  survey,'' \emph{Journal of Machine Learning Research}, vol.~20, no.~55, pp.
  1--21, 2019.

\bibitem{kingma_adam_2014}
D.~P. Kingma and J.~Ba, ``Adam: A method for stochastic optimization,''
  \emph{{arXiv} preprint {arXiv}:1412.6980}, 2014.

\bibitem{kwon:2020:deep_generative_graphs}
O.~Kwon and K.~Ma, ``\BIBforeignlanguage{English (US)}{A deep generative model
  for graph layout},'' \emph{\BIBforeignlanguage{English (US)}{IEEE
  Transactions on Visualization and Computer Graphics}}, vol.~26, no.~1, pp.
  665--675, 1 2020.

\bibitem{migut:2015:decision_boundaries}
M.~A. Migut, M.~Worring, and C.~J. Veenman, ``Visualizing multi-dimensional
  decision boundaries in {2D},'' \emph{Data Mining and Knowledge Discovery},
  vol.~29, no.~1, pp. 273--295, Jan. 2015.

\bibitem{Hamel:2006:decision_boundaries_of_svm}
L.~{Hamel}, ``Visualization of support vector machines with unsupervised
  learning,'' in \emph{2006 IEEE Symposium on Computational Intelligence and
  Bioinformatics and Computational Biology}, 2006, pp. 1--8.

\bibitem{schulz:2015:decision_boundaries}
A.~Schulz, A.~Gisbrecht, and B.~Hammer, ``Using {Discriminative}
  {Dimensionality} {Reduction} to {Visualize} {Classifiers},'' \emph{Neural
  Processing Letters}, vol.~42, no.~1, pp. 27--54, Aug. 2015.

\bibitem{bengio:2013:deep}
Y.~Bengio, ``Deep learning of representations: Looking forward,'' in
  \emph{International Conference on Statistical Language and Speech
  Processing}.\hskip 1em plus 0.5em minus 0.4em\relax Springer, 2013, pp.
  1--37.

\bibitem{chollet:2015:keras}
F.~Chollet \emph{et~al.}, ``Keras,'' \url{https://keras.io}, 2015.

\bibitem{google:2019:kerastuner}
G.~LLC, ``Keras-tuner,'' \url{https://github.com/keras-team/keras-tuner}, 2019.

\bibitem{abadi:2015:tensorflow}
M.~Abadi, A.~Agarwal, P.~Barham, E.~Brevdo, Z.~Chen, C.~Citro, G.~S. Corrado,
  A.~Davis, J.~Dean, M.~Devin, S.~Ghemawat, I.~Goodfellow, A.~Harp, G.~Irving,
  M.~Isard, Y.~Jia, R.~Jozefowicz, L.~Kaiser, M.~Kudlur, J.~Levenberg,
  D.~Man\'{e}, R.~Monga, S.~Moore, D.~Murray, C.~Olah, M.~Schuster, J.~Shlens,
  B.~Steiner, I.~Sutskever, K.~Talwar, P.~Tucker, V.~Vanhoucke, V.~Vasudevan,
  F.~Vi\'{e}gas, O.~Vinyals, P.~Warden, M.~Wattenberg, M.~Wicke, Y.~Yu, and
  X.~Zheng, ``{TensorFlow}: Large-scale machine learning on heterogeneous
  systems,'' 2015.

\bibitem{paszke:2019:pytorch}
A.~Paszke, S.~Gross, F.~Massa, A.~Lerer, J.~Bradbury, G.~Chanan, T.~Killeen,
  Z.~Lin, N.~Gimelshein, L.~Antiga, A.~Desmaison, A.~Kopf, E.~Yang, Z.~DeVito,
  M.~Raison, A.~Tejani, S.~Chilamkurthy, B.~Steiner, L.~Fang, J.~Bai, and
  S.~Chintala, ``Pytorch: An imperative style, high-performance deep learning
  library,'' in \emph{Advances in Neural Information Processing Systems 32},
  H.~Wallach, H.~Larochelle, A.~Beygelzimer, F.~d\textquotesingle
  Alch\'{e}-Buc, E.~Fox, and R.~Garnett, Eds.\hskip 1em plus 0.5em minus
  0.4em\relax Curran Associates, Inc., 2019, pp. 8024--8035.

\end{thebibliography}
%

\begin{IEEEbiography}[{\includegraphics[width=1in,height=1.25in,clip,keepaspectratio]{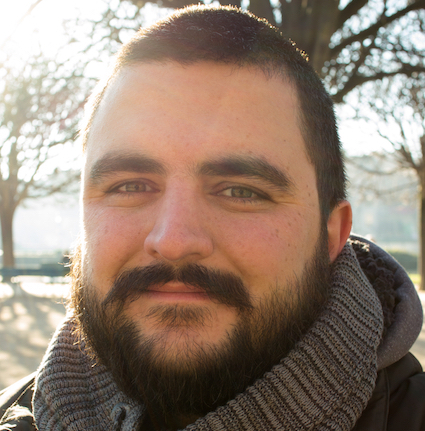}}]{Mateus Espadoto} received his PhD in Computer Science from the Institute of Mathematics and Statistics, University of São Paulo and from the Bernoulli Institute, University of Groningen. He has about 20 years of experience in data science and software development. His research interests include machine learning, high-dimensional data visualization and visual analytics.
\end{IEEEbiography}

\begin{IEEEbiography}[{\includegraphics[width=1in,height=1.25in,clip,keepaspectratio]{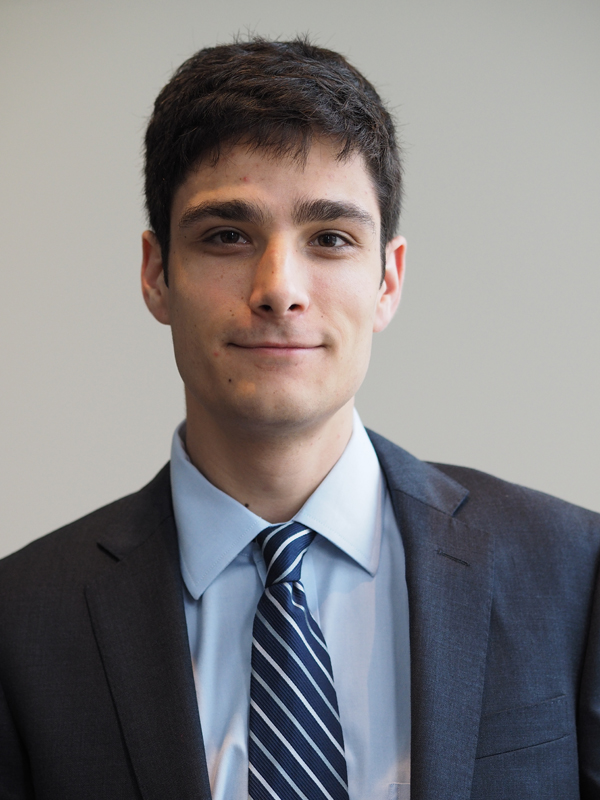}}]{Gabriel Appleby} received his MA degree in Computer Science at Tufts University where he is currently working towards his PhD. His research spans the fields of data visualization, visual analytics, and machine learning.
\end{IEEEbiography}

\begin{IEEEbiography}[{\includegraphics[width=1in,height=1.25in,clip,keepaspectratio]{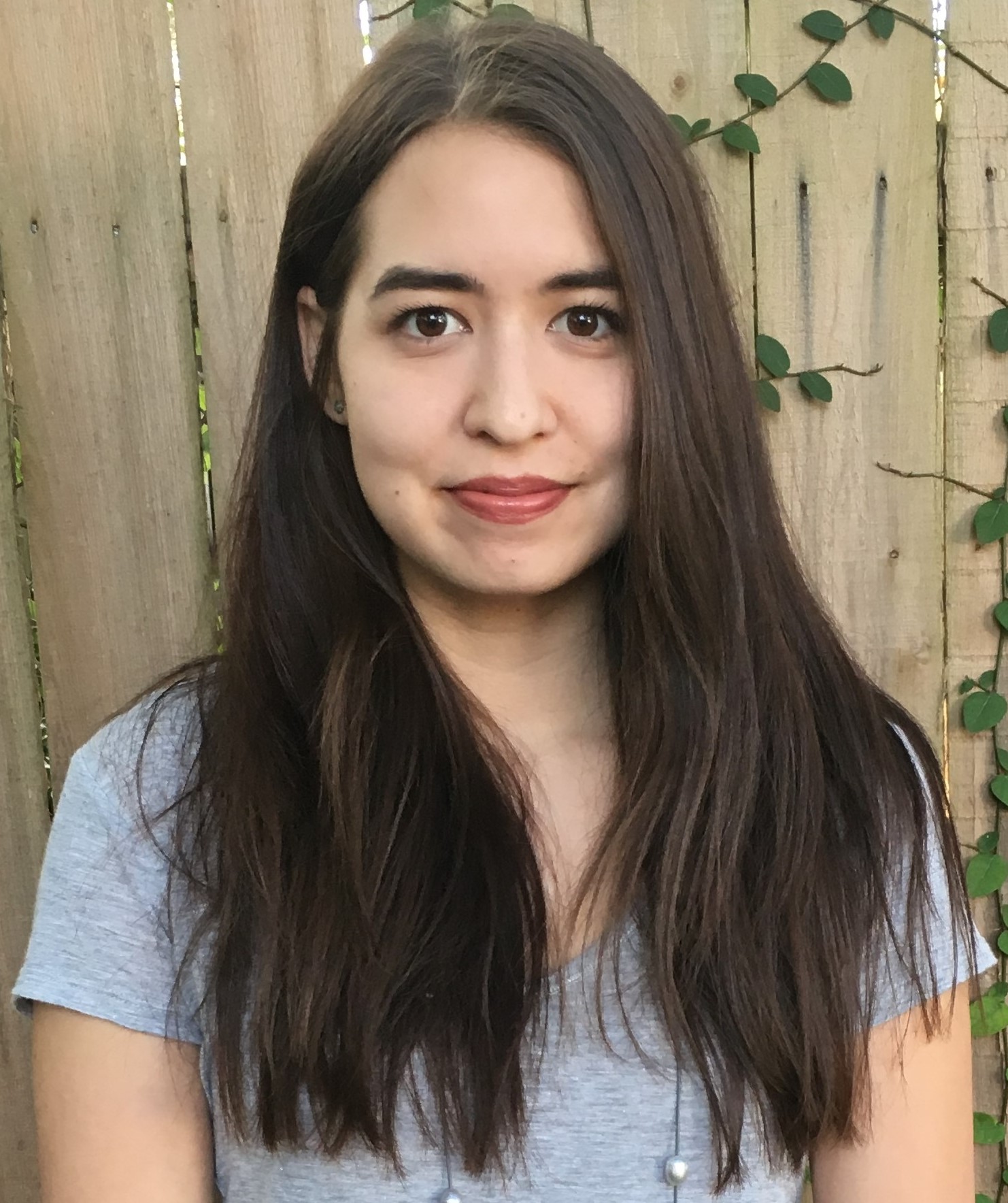}}]{Ashley Suh}
received her MS in Computer Science from Tufts University where she is currently pursuing a PhD. Her research interests include information visualization, visual analytics, and graph visualization.
\end{IEEEbiography}

\begin{IEEEbiography}[{\includegraphics[width=1in,height=1.25in,clip,keepaspectratio]{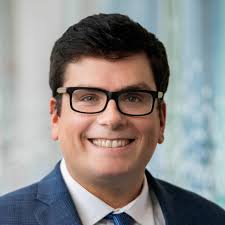}}]{Dylan Cashman} received his PhD in Computer Science from Tufts University. He received a bachelor of science in Mathematics from Brown University. Since 2020 he is a senior expert in data science and advanced visual analytics in the Insights, Strategies, and Design group at Novartis Pharmaceuticals. His research interests include visualization for data science and interactive machine learning.
\end{IEEEbiography}

\begin{IEEEbiography}[{\includegraphics[width=1in,height=1.25in,clip,keepaspectratio]{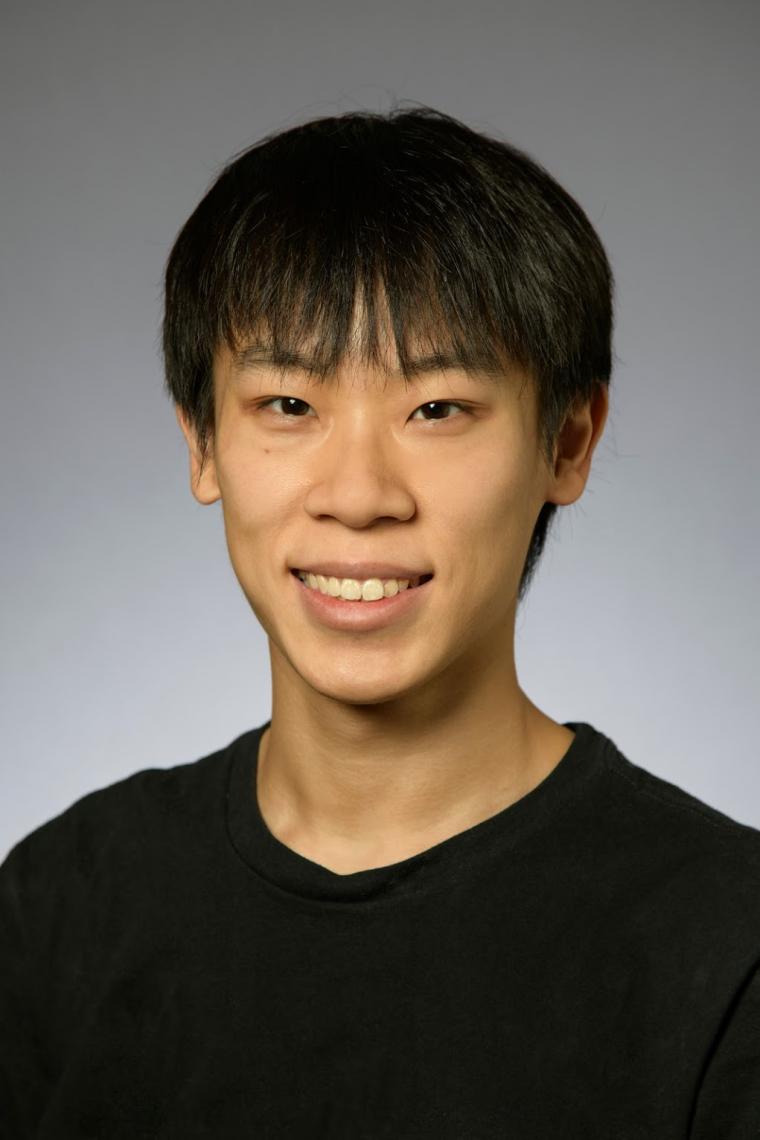}}]{Mingwei Li} received his PhD in Computer Science from University of Arizona. He received the BEng in electronics engineering from the Hong Kong University of Science and Technology. His research interests include data visualization and machine learning.
\end{IEEEbiography}

\begin{IEEEbiography}[{\includegraphics[width=1in,height=1.25in,clip,keepaspectratio]{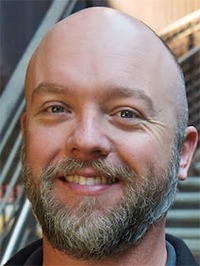}}]{Carlos Scheidegger}
Carlos Scheidegger received his PhD from the University of Utah, where he worked on software infrastructure for scientific collaboration. He is an assistant professor in the Department of Computer Science, University of Arizona. His research interests include large-scale data analysis, information visualization, and more broadly, what happens when people meet data.
\end{IEEEbiography}

\begin{IEEEbiography}[{\includegraphics[width=1in,height=1.25in,clip,keepaspectratio]{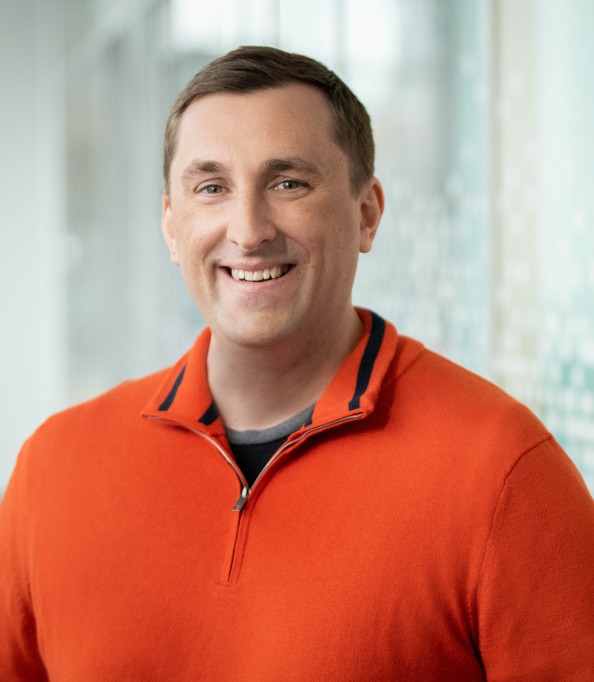}}]{Erik W Anderson}
received his PhD (2011) in Scientific Computing from the University of Utah, USA.  He was a senior scientist for Electrical Geodesics, Inc (EGI) until 2017 and then worked in research and development at Philips Neuro until 2020. Since 2020 he has been the Head of the Visualization and Visual Analytics group and Novartis Inc’s AI Innovation Center in Cambridge, MA USA.  His interests include high-dimensional visualization, multi-modal modeling and visualization, and biomedical image visualization
\end{IEEEbiography}

\begin{IEEEbiography}[{\includegraphics[width=1in,height=1.25in,clip,keepaspectratio]{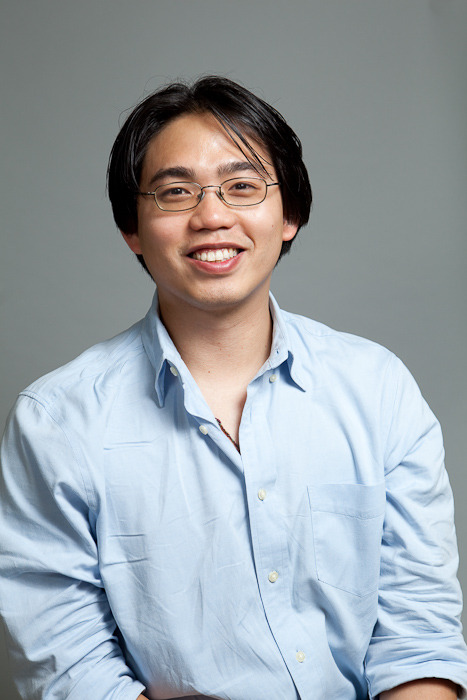}}]{Remco Chang} received his PhD in computer science from the University of North Carolina Charlotte. He is an associate professor in computer science with Tufts University. His research interests include visual analytics, information visualization, human computer interaction, and databases.
\end{IEEEbiography}

\begin{IEEEbiography}[{\includegraphics[width=1in,height=1.25in,clip,keepaspectratio]{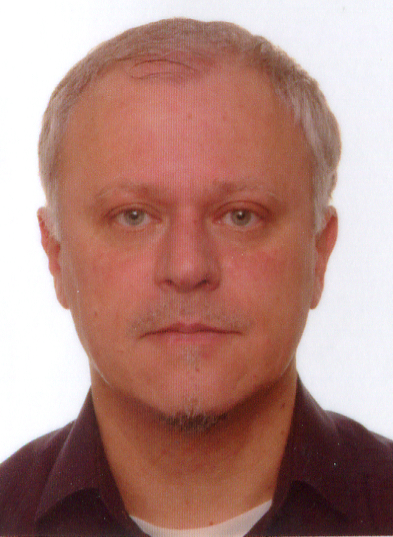}}]{Alexandru C. Telea} received his PhD (2000) in Computer Science from the Eindhoven University of Technology, the Netherlands. He was assistant professor in visualization and computer graphics at the same university (until 2007) and then full professor of visualization at the University of Groningen, the Netherlands. Since 2019 he is full professor of visual data analytics at Utrecht University, the Netherlands. His interests include high-dimensional visualization, visual analytics, and image-based information visualization.
\end{IEEEbiography}




\end{document}


\setcounter{section}{7}
\onecolumn
\section{Supplemental}
\label{sec:supplemental}
\begin{small}
\begin{longtable}{|r|r|r|r|r|r|r|r|}
\toprule
 Layer one &  Layer two &  Layer three &  Layer four &  Dropout &       MAE &       STD &      N \\
\midrule
\endhead
\midrule
\multicolumn{8}{r}{{Continued on next page}} \\
\midrule
\endfoot

\bottomrule
\endlastfoot
        64 &        128 &          256 &         512 &   0.0000 &  0.036941 &  0.000003 &  42000 \\
       128 &        256 &          512 &        1024 &   0.0000 &  0.036944 &  0.000025 &  42000 \\
       256 &        512 &         1024 &        2048 &   0.0000 &  0.036945 &  0.000021 &  42000 \\
       640 &       1280 &         1280 &         640 &   0.0000 &  0.036950 &  0.000033 &  42000 \\
       240 &        240 &          240 &         240 &   0.0000 &  0.036961 &  0.000031 &  42000 \\
       480 &        480 &          480 &         480 &   0.0000 &  0.036966 &  0.000009 &  42000 \\
       640 &        320 &          320 &         640 &   0.0000 &  0.036966 &  0.000013 &  42000 \\
       160 &        320 &          320 &         160 &   0.0000 &  0.036973 &  0.000021 &  42000 \\
\end{longtable}
\end{small}
\begin{small}
\begin{longtable}{|r|r|r|r|r|r|r|r|}
\toprule
 Layer one &  Layer two &  Layer three &  Layer four &  Dropout &       MAE &       STD &      N \\
\midrule
\endhead
\midrule
\multicolumn{8}{r}{{Continued on next page}} \\
\midrule
\endfoot

\bottomrule
\endlastfoot
       128 &        256 &          512 &        1024 &   0.0000 &  0.062410 &  0.000425 &  42000 \\
       640 &        320 &          320 &         640 &   0.0625 &  0.062606 &  0.000113 &  42000 \\
       640 &        320 &          320 &         640 &   0.0000 &  0.062787 &  0.000551 &  42000 \\
       480 &        480 &          480 &         480 &   0.0625 &  0.063030 &  0.000050 &  42000 \\
       480 &        480 &          480 &         480 &   0.0000 &  0.063168 &  0.000258 &  42000 \\
       256 &        512 &         1024 &        2048 &   0.0000 &  0.063341 &  0.000206 &  21000 \\
       320 &        640 &          640 &         320 &   0.0625 &  0.063459 &  0.000019 &  42000 \\
       320 &        640 &          640 &         320 &   0.0000 &  0.063574 &  0.000372 &  42000 \\
\end{longtable}
\end{small}
\begin{small}
\begin{longtable}{|r|r|r|r|r|r|r|r|}
\toprule
 Layer one &  Layer two &  Layer three &  Layer four &  Dropout &       MAE &       STD &      N \\
\midrule
\endhead
\midrule
\multicolumn{8}{r}{{Continued on next page}} \\
\midrule
\endfoot

\bottomrule
\endlastfoot
      1024 &       2048 &         4096 &        8192 &   0.0000 &  0.072804 &  0.000411 &  42000 \\
      1280 &       2560 &         2560 &        1280 &   0.0000 &  0.072873 &  0.000136 &  42000 \\
       512 &       1024 &         2048 &        4096 &   0.0000 &  0.073108 &  0.000268 &  42000 \\
      1280 &        640 &          640 &        1280 &   0.0000 &  0.073209 &  0.000065 &  42000 \\
       256 &        512 &         1024 &        2048 &   0.0000 &  0.073214 &  0.000640 &  42000 \\
       960 &        960 &          960 &         960 &   0.0000 &  0.073279 &  0.000274 &  42000 \\
      1920 &       1920 &         1920 &        1920 &   0.0000 &  0.073541 &  0.000657 &  42000 \\
       128 &        256 &          512 &        1024 &   0.0000 &  0.073631 &  0.000554 &  42000 \\
\end{longtable}
\end{small}

\begin{small} 
\begin{longtable}{|r|r|r|r|r|r|r|r|}
\toprule
 Layer one &  Layer two &  Layer three &  Layer four &  Dropout &       MAE &       STD &      N \\
\midrule
\endhead
\midrule
\multicolumn{8}{r}{{Continued on next page}} \\
\midrule
\endfoot

\bottomrule
\endlastfoot
        64 &        128 &          256 &         512 &   0.0000 &  0.011698 &  0.000489 &  42000 \\
       256 &        512 &         1024 &        2048 &   0.0000 &  0.012288 &  0.001286 &  42000 \\
       640 &       1280 &         1280 &         640 &   0.0000 &  0.013136 &  0.000805 &  42000 \\
       160 &        320 &          320 &         160 &   0.0000 &  0.013209 &  0.001000 &  42000 \\
       640 &        320 &          320 &         640 &   0.0000 &  0.013929 &  0.001523 &  42000 \\
       128 &        256 &          512 &        1024 &   0.0000 &  0.013969 &  0.000097 &  42000 \\
       320 &        640 &          640 &         320 &   0.0000 &  0.014385 &  0.001054 &  42000 \\
      1280 &        640 &          640 &        1280 &   0.0000 &  0.014422 &  0.000853 &  42000 \\
\end{longtable}
\end{small}